\documentclass[twocolumn]{aastex631}

\renewcommand{\today}{December 5, 2025}

\usepackage{graphicx}
\usepackage{mathtools} 
\usepackage{xcolor}
\usepackage{xspace}
\usepackage{hyperref}
\usepackage[version=4,arrows=pgf-filled]{mhchem}

\def\NHth{NH$_3$\xspace}
\def\NHtD{NH$_2$D\xspace}
\def\Nt{N$_2$\xspace}
\def\NtHp{N$_2$H$^+$\xspace}
\def\NtDp{N$_2$D$^+$\xspace}

\def\DCOp{DCO$^+$\xspace}

\def\Ht{H$_2$\xspace}

\def\Hthp{H$_3^+$\xspace}
\def\HtDp{H$_2$D$^+$\xspace}
\def\oHtDp{ortho-H$_2$D$^+$\xspace}
\def\oHtDpgrd{1$_{10}$--1$_{11}$\xspace}
\def\DtHp{D$_2$H$^+$\xspace}
\def\pDtHp{para-D$_2$H$^+$\xspace}
\def\pDtHpgrd{1$_{10}$--1$_{01}$\xspace}
\def\Dtp{D$_3^+$\xspace}

\def\mJybm{mJy\,beam$^{-1}$\xspace}
\def\mJybmkms{mJy\,beam$^{-1}$\,km\,s$^{-1}$\xspace}
\def\kms{km\,s$^{-1}$\xspace}
\def\arcmin{\mbox{$^{\prime}$}\xspace}
\def\arcsec{\mbox{$^{\prime\prime}$}\xspace}
\def\soufullname{G205.46$-$14.56\,M3\xspace}
\def\souname{G205\,M3\xspace}

\renewcommand\micron{\mbox{$\mu$m}\xspace}

\accepted{September 25, 2025}

\begin{document}

\title{
Unveiling Central \oHtDp Depletion at Sub-kau Scales in Prestellar Core \soufullname: 
The First Interferometric Evidence and Implications for Deuterium Chemistry}

\author[0000-0002-6868-4483]{Sheng-Jun Lin}
\affiliation{Academia Sinica Institute of Astronomy and Astrophysics, No. 1, Section 4, Roosevelt Road, Taipei 10617, Taiwan}
\email{shengjunlin@asiaa.sinica.edu.tw}

\author[0000-0003-4603-7119]{Sheng-Yaun Liu}
\affiliation{Academia Sinica Institute of Astronomy and Astrophysics, No. 1, Section 4, Roosevelt Road, Taipei 10617, Taiwan}
\email{syliu@asiaa.sinica.edu.tw}

\author[0000-0002-4393-3463]{Dipen Sahu}
\affiliation{Academia Sinica Institute of Astronomy and Astrophysics, No. 1, Section 4, Roosevelt Road, Taipei 10617, Taiwan}
\affiliation{Physical Research Laboratory, Navrangpura, Ahmedabad, Gujarat 380009, India}

\author[0000-0002-3319-1021]{Laurent Pagani}
\affiliation{LUX, Observatoire de Paris, PSL University, Sorbonne Universit\'es, CNRS, F-75014 Paris, France}

\author[0000-0002-3319-1021]{Tien-Hao Hsieh}
\affiliation{Academia Sinica Institute of Astronomy and Astrophysics, No. 1, Section 4, Roosevelt Road, Taipei 10617, Taiwan}
\affiliation{Taiwan Astronomical Research Alliance (TARA), Taiwan}

\author[0000-0001-9304-7884]{Naomi Hirano}
\affiliation{Academia Sinica Institute of Astronomy and Astrophysics, No. 1, Section 4, Roosevelt Road, Taipei 10617, Taiwan}

\author[0000-0001-5522-486X]{Shih-Ping Lai}
\affiliation{Institute of Astronomy, National Tsing Hua University, No. 101, Section 2, Kuang-Fu Road, Hsinchu 30013, Taiwan}
\affiliation{Center for Informatics and Computation in Astronomy (CICA), NTHU, No. 101, Section 2, Kuang-Fu Road, Hsinchu 30013, Taiwan}

\author[0000-0002-5286-2564]{Tie Liu}
\affiliation{Shanghai Astronomical Observatory, Chinese Academy of Sciences, 80 Nandan Road, Shanghai 200030, People's Republic of China}

\author[0000-0002-1369-1563]{Shih-Ying Hsu}
\affiliation{Academia Sinica Institute of Astronomy and Astrophysics, No. 1, Section 4, Roosevelt Road, Taipei 10617, Taiwan}

\author[0000-0003-1275-5251]{Shanghuo Li}
\affiliation{School of Astronomy and Space Science, Nanjing University, 163 Xianlin Avenue, Nanjing 210023, People’s Republic of China}

\author[0000-0003-2412-7092]{Kee-Tae Kim}
\affiliation{Korea Astronomy and Space Science Institute (KASI), 776 Daedeokdae-ro, Yuseong-gu, Daejeon 34055, Republic of Korea}
\affiliation{University of Science and Technology, Korea (UST), 217 Gajeong-ro, Yuseong-gu, Daejeon 34113, Republic of Korea}


\begin{abstract}
Prestellar cores represent the initial conditions of star formation,
but heavy molecules such as CO are strongly depleted in their cold, dense interiors,
limiting the ability to probe core centers. 
Deuterated molecular ions therefore emerge as key tracers 
because deuterium fractionation is enhanced at low temperatures. 
We present the first direct observation of \oHtDp depletion in the prestellar core \soufullname using ALMA 820~\micron continuum and \oHtDp(\oHtDpgrd) data at $\sim$300-au resolution. 
We confirm the previously reported two substructures, B1 and B2, and identify a central \oHtDp depletion zone toward B1 with $\sim$6$\sigma$ contrast and an inferred diameter $\lesssim$600~au, together with a peak $x(\text{\NtDp})/x(\text{\NtHp})=1.03^{+0.07}_{-0.56}$. 
The observationally inferred profiles of $x(\text{\oHtDp})$ and $x(\text{\NtDp})/x(\text{\NtHp})$ are reproduced by a deuteration-focused chemodynamical model; however, the central o-\HtDp depletion is only marginally matched within the $2\sigma$ upper limit, likely suggesting additional deuteration in the depletion zone.
From these models we infer a core age of $\sim$0.42~Ma, comparable to the freefall time, suggesting that the substructures formed via rapid, turbulence-dominated fragmentation rather than slow, quasistatic contraction.
Our observations also reveal that \oHtDp velocity dispersions are largely subsonic in the core
and nearly thermal between B1 and B2, consistent with turbulence dissipating within a few freefall times.
These results highlight the critical role of deuterated ions for both chemical evolution and dynamics in dense cores.
\end{abstract}


\section{Introduction} \label{sec:intro}

Starless cores represent the earliest phase of star and planet formation, with prestellar cores as a subset that are gravitationally unstable and on the verge of collapse \citep{diFrancesco07, Ward-Thompson07}. 
These prestellar cores, characterized by modest densities and low temperatures \citep[$\gtrsim10^5$~cm$^{-3}$ and $\lesssim 10$~K;][]{Keto08}, pose significant observational challenges.

One of the main difficulties to study the prestellar cores is that extensive molecular depletion causes a lack of tracers toward the dense centers.
Different depletion mechanisms operate in prestellar cores. 
Neutral molecules (e.g., CO) freeze directly onto dust grains, 
whereas molecular ions (e.g., HCO$^+$, \DCOp) become underabundant because their formation is suppressed by the loss of parent neutrals \citep[e.g.,][]{Pagani05, Bergin07}.

Nitrogen-bearing species (e.g., \NHth, \NtHp, \NtDp), being less affected by depletion \citep[e.g.,][]{Tafalla04, Pagani07, Redaelli19, Lin20}, have long served as reliable tracers of prestellar cores. However,
recent two high-resolution interferometric studies have resolved the depletion zones of the \NHth isotopologues within these prestellar cores.
The Atacama Large Millimeter/submillimeter Array (ALMA) observations by \citet{Caselli22} revealed complete depletion of \NHtD over a 3,600 au diameter in L1544,
while the Very Large Array (VLA) observations by \citet{Pineda22} reported \NHth depletion in Oph-H-MM1 within a $\sim$4,000 au-diameter zone with $n_{\rm H_2}\geq 2\times10^5$~cm$^{-3}$.

Once the heavy species are depleted in the dense center, 
\Hthp becomes the dominant molecular ion. 
Then, deuterium fractionation is significantly enhanced in the cold innermost region \citep{Flower04b, Walmsley04, Pagani09b}.
Consequently, the \Hthp isotopologues, including \HtDp, \DtHp, and \Dtp, become abundant.
These ions can further boost molecular deuteration by reacting with residual, partially depleted molecules (e.g., \Nt) to form deuterated ions (e.g., \NtDp).

Since \HtDp represents the first deuteration step of \Hthp, its abundance is immediately enhanced following the depletion of heavy species.
However, as core densities rise, additional chemical processes decrease its abundance either by converting \HtDp into \DtHp and \Dtp through enhanced fractionation, or by re-hydrogenation as temperature rises once a protostar forms. Moreover, their parent molecule HD, the primary deuterium reservoir, may be depleted as deuterium is diverted to HDO on grain surfaces rather than HD \citep{Sipila13}. This diversion reduces the deuteration of \Ht, \Hthp, and \NtHp isotopologues, although confirming this pathway remains challenging due to difficulties in detecting HDO on grains \citep{Slavicinska24}.

In prestellar cores, the only accessible transitions of \Hthp isotopologues are the ground transitions of \oHtDp(\oHtDpgrd) at 372~GHz, and \pDtHp(\pDtHpgrd) at the challenging 691~GHz.
The others are either more difficult to observe (p-\HtDp and o-\DtHp, in the THz regime)
or unavailable (\Hthp and \Dtp, due to their lack of a dipole moment).
While o-\HtDp has been detected in several nearby prestellar cores with single-dish observations
\citep[e.g.,][]{Caselli08, Pagani09b, Lin20, Lin24},
to date the interferometric \oHtDp maps toward prestellar cores have only been presented by \citet{Friesen24}, observing the low-mass Oph-A-SM1N core, and by \citet{Redaelli21}, targeting two high-mass prestellar clumps, with o-\HtDp emission peaking at the continuum centers.
A recent interferometric search for o-\HtDp toward L1544 instead resulted in a nondetection, suggesting an interferometrically resolved-out extended distribution  \citep{Tokuda25}.
However, sensitivity limits might also contribute to the apparent nondetection of potentially more compact structures \citep[e.g.,][]{Caselli19}.

The first \oHtDp depletion detection was reported by \citet{Pagani24}, who presented the simultaneous JCMT and APEX single-dish maps of both \oHtDp(\oHtDpgrd) and \pDtHp(\pDtHpgrd) toward IRAS\,16293E at 14\arcsec resolution (1,974~au at a distance of 141~pc).
Their study found a marginal \oHtDp depletion (2.3$\sigma$), 
coinciding with a central \pDtHp peak,
consistent with further \Hthp deuteration.
Given that deuterium fractionation has been widely used as a chemical clock, this observation highlights its value as a core evolution indicator \citep{Parise11, Pagani13, Bovino21}.
In contrast, the earlier claim of interferometric \oHtDp depletion in Oph-A-SM1N \citep{Friesen14} was later attributed to poor data quality.

Recently, the ALMA Survey of Orion Planck Galactic Cold Clumps \citep[ALMASOP;][]{Dutta20} identified a unique prestellar core,  \soufullname (hereafter \souname), located at a distance of $\sim$400~pc \citep{Kounkel17, Zucker19}. This core is extremely centrally concentrated with a peak density of $1.1\times 10^7$~cm$^{-3}$, and
harbors two substructures (B1 and B2) separated by 1,200~au, potentially indicative of binary formation \citep{Sahu21, Sahu23}.
Such a dense star-forming environment is the ideal target for probing deuterated species.

In this paper, we present the first interferometric discovery of an \oHtDp (hereafter o-\HtDp) depletion zone in \souname, which we attribute to enhanced deuteration. 
This interpretation is supported by our \NtHp(1--0 and 4--3) data together with archival \NtDp(3--2) observations.
Section~\ref{sec:obs} describes the observations. 
Section~\ref{sec:res} presents the results. 
Section~\ref{sec:ana} details the analysis, including LTE/non-LTE radiative transfer, kinematics, and chemodynamical modeling.
Section~\ref{sec:dis} discusses the chemical implications, and Section~\ref{sec:con} summarizes the conclusions.

\section{Observations} \label{sec:obs}

\begin{deluxetable*}{lrccccccc}
\caption{Summary of the Data Parameters}
\tablehead{
\colhead{Line/Continuum} & \colhead{$\nu_0$} & \colhead{Array Config.} & \colhead{$\theta_{\rm res}$} & \colhead{$\theta_{\rm MRS}$} & \colhead{$\delta v_{\rm ch}$} & \colhead{$\delta v_{\rm res}$} & \multicolumn{2}{c}{rms Noise Level}\\
\colhead{} & \colhead{(GHz)} & \colhead{} & \colhead{(arcsec., degree)} & \colhead{(arcsec.)} & \colhead{(km s$^{-1}$)} &  \colhead{(km s$^{-1}$)} & \colhead{(mJy beam$^{-1}$)} & \colhead{(K)}
}
\colnumbers
\startdata
\multicolumn{9}{c}{\textit{ALMA Data}}\\
820\,\micron continuum & 357--373      & 7\,m, C-1       & 0\farcs90$\times$0\farcs77, $-$73\fdg7 & 18\arcsec  & ...   & ...   & 0.5 & $7\times10^{-3}$ \\ 
o-\HtDp($J_{K_aK_c}$=\oHtDpgrd)      & 372.4213558   & 7\,m, C-1       & 0\farcs86$\times$0\farcs73, $-$68\fdg5 & 18\arcsec  & 0.098 & 0.114 & 12 & 0.17 \\ 
\NtHp($J_{F_1F}$=4$_{56}$--3$_{45}$)            & 372.6725370   & 7\,m, C-1       & 0\farcs87$\times$0\farcs73, $-$70\fdg8 & 18\arcsec  & 0.098 & 0.114 &  13 & 0.18 \\ 
\NtDp($J_{F_1F}$=3$_{45}$--2$_{34}$)            & 231.3219281   & TP, 7\,m, C-1   & 2\farcs41$\times$1\farcs32, $-$69\fdg6 & ...        & 0.040 & 0.046 &  39 & 0.28 \\ 
\NtHp($J_{F_1}$=1$_0$--0$_1$)            & 93.1762543    & TRAO, 7\,m, C-3       & 2\farcs10$\times$1\farcs59, $-$87\fdg0 & ...   & 0.049 & 0.098 & 5.0 & 0.21 \\ 
\multicolumn{9}{c}{\textit{JCMT Data}}\\
850\,\micron continuum & 353           &  ...            & 14\arcsec                              & ...        & ...   & ...   & 2.6 & $1.3\times10^{-4}$ 
\enddata
\tablecomments{Columns (1)--(2): rest frequencies of the spectral lines are taken from \citet{Jusko17} for \HtDp and from \citet{Pagani09a} for \NtHp isotopologues. 
Column (3): array configurations used for image combination. 
Column (4): synthesized beam or single-dish beam. 
Column (5): maximum recoverable scale for the ALMA ACA array, based on the fifth percentile baseline length, for observations without single-dish data.
Column (6): native channel spacing in data. 
Column (7): spectral resolution (see details in \autoref{app:other_data}). 
Columns (8)--(9): rms noise level, provided in intensity and also brightness temperature assuming Gaussian beams of $\theta_{\rm res}$.\label{tab:obs}}
\end{deluxetable*}

\subsection{ALMA Band 7 and Band 3 Observations}
Our ALMA Band 7 observation of \souname was carried out in Cycle-9 under project 2022.1.01603.S (PI: Sheng-Jun Lin). Observations were conducted in two array configurations: 
the 12\,m array C-1 configuration, and 7\,m array of the Atacama Compact Array (ACA). 
Their projected baseline ranges were 17.4--381~$k\lambda$, and 10.6--55.7~$k\lambda$, respectively.
The phase calibrator was either J0552+0313 or J0541$-$0541, while the flux and bandpass calibrator was J0423$-$0120.
Our ALMA Band 3 observations of \souname were conducted in Cycle-8 under project 2021.1.00546.S (PI: Dipen Sahu), utilizing three array configurations: 
the 12\,m array C-7 and C-3 configurations, and 7\,m array of the ACA.
Their projected baseline ranges were 4.49--960~$k\lambda$, 4.38--152~$k\lambda$, and 2.45--13.7~$k\lambda$, respectively.
Phase calibration was performed using J0552+0313 (12\,m) and J0542$-$0913 (7\,m), and flux/bandpass calibration employed J0510+1800 (12\,m) and J0538$-$4405 (7\,m).
The phase center coordinates in both bands were 
$\alpha_{\rm ICRS}=5^{\mathrm h}46^{\mathrm m}05\fs960$, $\delta_{\rm ICRS}=-0\degr09\arcmin32\farcs45$.

\begin{figure*}[th]
    \centering
    \includegraphics[width=\textwidth]{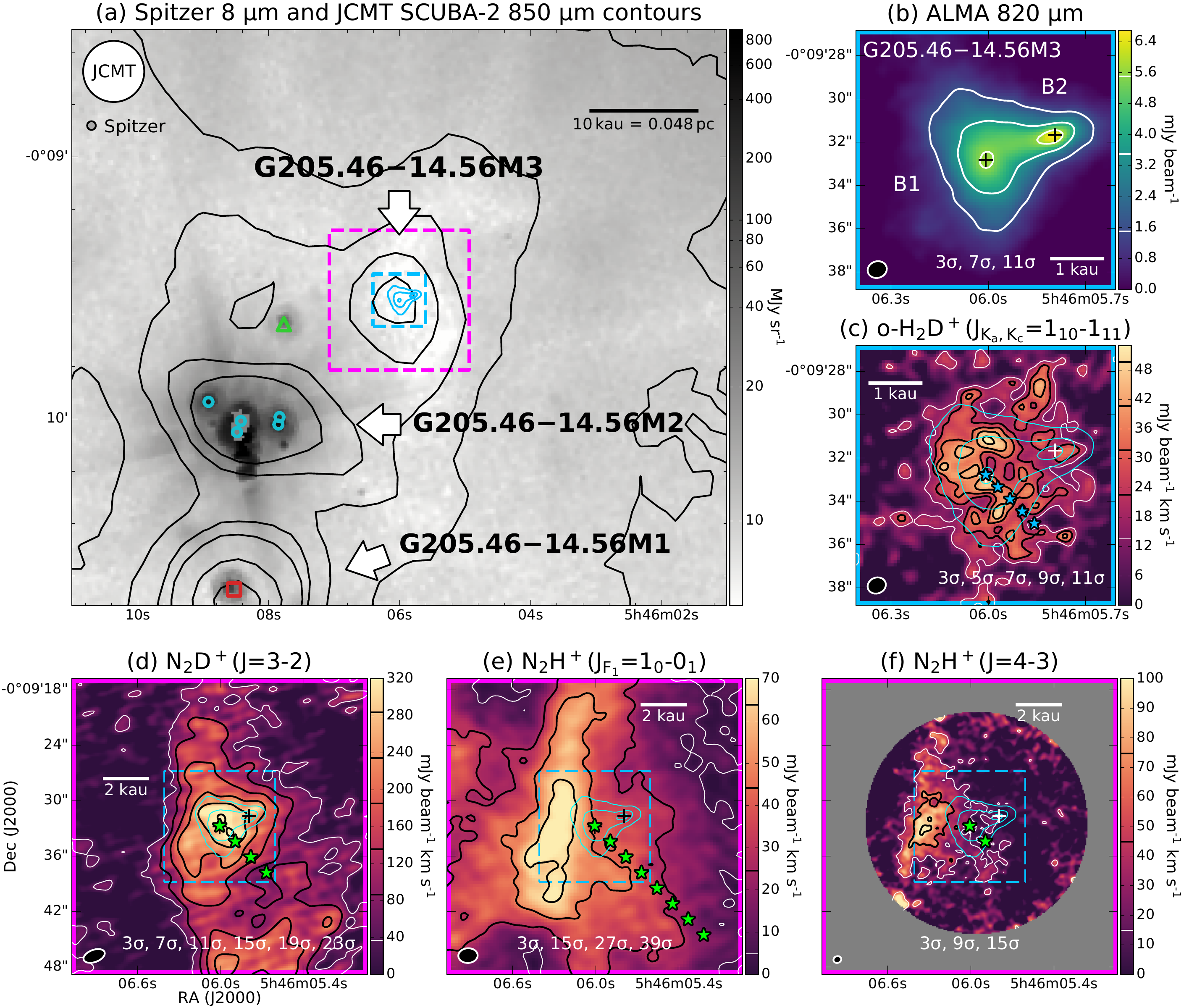}
    \caption{
    Multiscale view of the prestellar core \soufullname in dust continuum and integrated molecular line emission. 
    \textbf{(a)} \textit{Spitzer} IRAC 8~\micron image of the three JCMT dense cores G205.46$-$14.56 M1, M2, and M3 \citep{Dutta20}.
    JCMT SCUBA-2 850~\micron continuum contours overlaid at 10--90\% (20\% steps) of the peak of the faintest prestellar core, \souname (500~\mJybm), with additional contours at 1000 and 2000~\mJybm.
    Colored symbols mark protostars compiled by \citet{Reipurth23}:
    five cyan circles denote the Class 0/I multiple system SSV\,63 hosted by M2,  
    one red square marks the Class 0 source HOPS\,317 embedded in M1,  
    and the green triangle indicates the Class II source IRS\,1.
    A cyan dashed box outlines the region enlarged in panels~(b) and (c), and a magenta box encloses panels~(d)--(f); these same borders are repeated around the zoom-in panels to guide the eye.
    \textbf{(b)} ALMA Band\,7 820~\micron continuum. 
    \textbf{(c)} ALMA Band\,7 \oHtDp(\oHtDpgrd) map integrated over $v_{\rm LSR}=9.5$--$10.8$~\kms.
    \textbf{(d)} ALMA Band\,6 \NtDp(3--2) map integrated over $v_{\rm LSR}=9.1$--$11.6$~\kms, where the Total-Power (TP) data are included to recover extended emission. 
    \textbf{(e)} ALMA Band\,3 \NtHp(1--0) isolated hyperfine component integrated over $v_{\rm LSR}=9.0$--$11.1$~\kms and combined with TRAO single-dish data.
    \textbf{(f)} ALMA Band\,7 \NtHp(4--3) map integrated over $v_{\rm LSR}=9.6$--$10.9$~\kms.
    ALMA emission contour levels are labeled in each panel; the $3\sigma$ contour is drawn in white to maximize contrast against the color scale, whereas higher levels are black for molecular lines.
    These 1$\sigma$ levels are 0.5~\mJybm in panel~(b), and 4.5, 12, 1.6, and 5~\mJybmkms in panels~(c)--(f), respectively.
    Contours from panel~(b) are overlaid on all other panels for reference. 
    Crosses denote the substructures B1 and B2 \citep{Sahu21}, asterisks mark positions adopted for the non-LTE radiative-transfer analysis (Section~\ref{sec:ana_nlte}). 
    Beam sizes and scale bars are indicated in each panel.
    We note that the ALMA color images are primary-beam corrected, 
    whereas contours are drawn from uncorrected images with uniform noise.}
    \label{fig:cont_mole}
\end{figure*}

Each band had four spectral windows (SPWs) in the correlator.
We focus on the Band 7 SPWs containing o-\HtDp(\oHtDpgrd), \NtHp(4--3), and continuum, and on the Band 3 SPW of \NtHp(1--0). The full correlator setup is provided in \autoref{app:other_data}.
The raw data were calibrated using the corresponding pipelines in CASA \citep{casa2022}: version 6.4.1 for Band 7 and version 6.2.1 for Band 3.
Subsequent imaging used CASA 6.4.1 with the \texttt{tclean} task.
Band 7 continuum data were obtained by averaging line-free channels, yielding an aggregate bandwidth of 2.29~GHz.
Line data were processed by subtracting the continuum in the visibility domain.
To better recover extended emission, 
we used the \texttt{multiscale} deconvolver with scales of 0, 1, and 3 beam sizes.
For Band 7, the data were imaged with a Briggs weighting scheme and the robustness of $+1.0$ to enhance the signal-to-noise ratio (S/N).
For Band 3, only the 12\,m C-3 and 7\,m data for the \NtHp(1--0) SPW were imaged with a Briggs \texttt{robust} parameter of $+0.5$, because \souname was resolved out in the C-7 configuration at a resolution of $\sim$0\farcs4.
Our \NtHp(1--0) data was further combined with archival Taeduk Radio Astronomy Observatory 14\,m single-dish data (see Section~\ref{sec:obs_arch}) using the \texttt{feather} task. 
We compared the integrated-intensity maps from the ALMA-only data and the \texttt{feather}ed data for the isolated hyperfine component $J_{F_1}=1_0-0_1$ of \NtHp.
We find that only roughly 30\% of the flux is resolved out by the ALMA-only observations toward \souname B1 within the core boundary \citep[10,720~au;][]{Sahu23}.
This implies that most of the emission arises on small scales, with a modest extended component.
This is also consistent with \souname being compact relative to other Orion prestellar cores that suffer stronger interferometric filtering \citep{Dutta20}.
The resulting 820~\micron continuum, o-\HtDp, and \NtHp(1--0) and (4--3) images are shown in \autoref{fig:cont_mole}
with spatial resolutions and maximum recoverable scales (MRSs) summarized in \autoref{tab:obs}.

\subsection{Archival Data}\label{sec:obs_arch}

We utilized archival ALMA Band 6 \NtDp(3--2) data observed in Cycle-4 under project 2016.1.01338.S (PI: D.~Mardones), as published by \citet{HsiehCH19}.
Their observations targeted a mosaic field covering \souname with the 12\,m array in the C-1 configuration, and ACA including both the 7\,m array and total-power (TP) array for large-scale emission.
To include large-scale emission in our ALMA \NtHp(1--0) data, we incorporated the Taeduk Radio Astronomy Observatory (TRAO) 14\,m single-dish \NtHp(1--0) on-the-fly mapping data 
with the beam size of 54\farcs1
published by \citet{Yoo23}.
We also retrieved the James Clerk Maxwell Telescope (JCMT) SCUBA-2 \citep{Holland13} 850~\micron data with the beam size of 14\arcsec from the Transient Survey \citep{Mairs24JCMT_Transient_Survey} under project M20AL007, spanning from Feb 2020 to Jan 2023, via the JCMT archive\footnote{http://www.cadc-ccda.hia-iha.nrc-cnrc.gc.ca/en/jcmt/}.
The data were reprocessed using the \texttt{skyloop} routine in Starlink \citep{Chapin13} with the \texttt{dimmconfig\_pca.lis} configuration optimized for extended emission and a flux conversion factor of 495~Jy~pW$^{-1}$ \citep{Mairs21}.
For further details on these observations, please refer to \citet{HsiehCH19, Yoo23, Mairs24JCMT_Transient_Survey}.
In addition, we retrieved \textit{Spitzer} IRAC-4 8~\micron mosaics (Level 2 products) with the beam size of 2\arcsec 
under program id 43 (PI: G. Fazio) from the Spitzer Heritage Archive (SHA)\footnote{https://sha.ipac.caltech.edu/applications/Spitzer/SHA/}.

\autoref{fig:cont_mole}(a) displays the JCMT 850~\micron continuum contours overlaid on the \textit{Spitzer} 8~\micron image, whereas \autoref{fig:cont_mole}(d) and \autoref{fig:cont_mole}(e) present the multi-array combined images of \NtDp(3--2) and \NtHp(1--0), respectively.
The observational parameters for the (sub-)millimeter data are listed in \autoref{tab:obs}.

\begin{figure*}[th]
    \centering
    \includegraphics[width=0.8\textwidth]{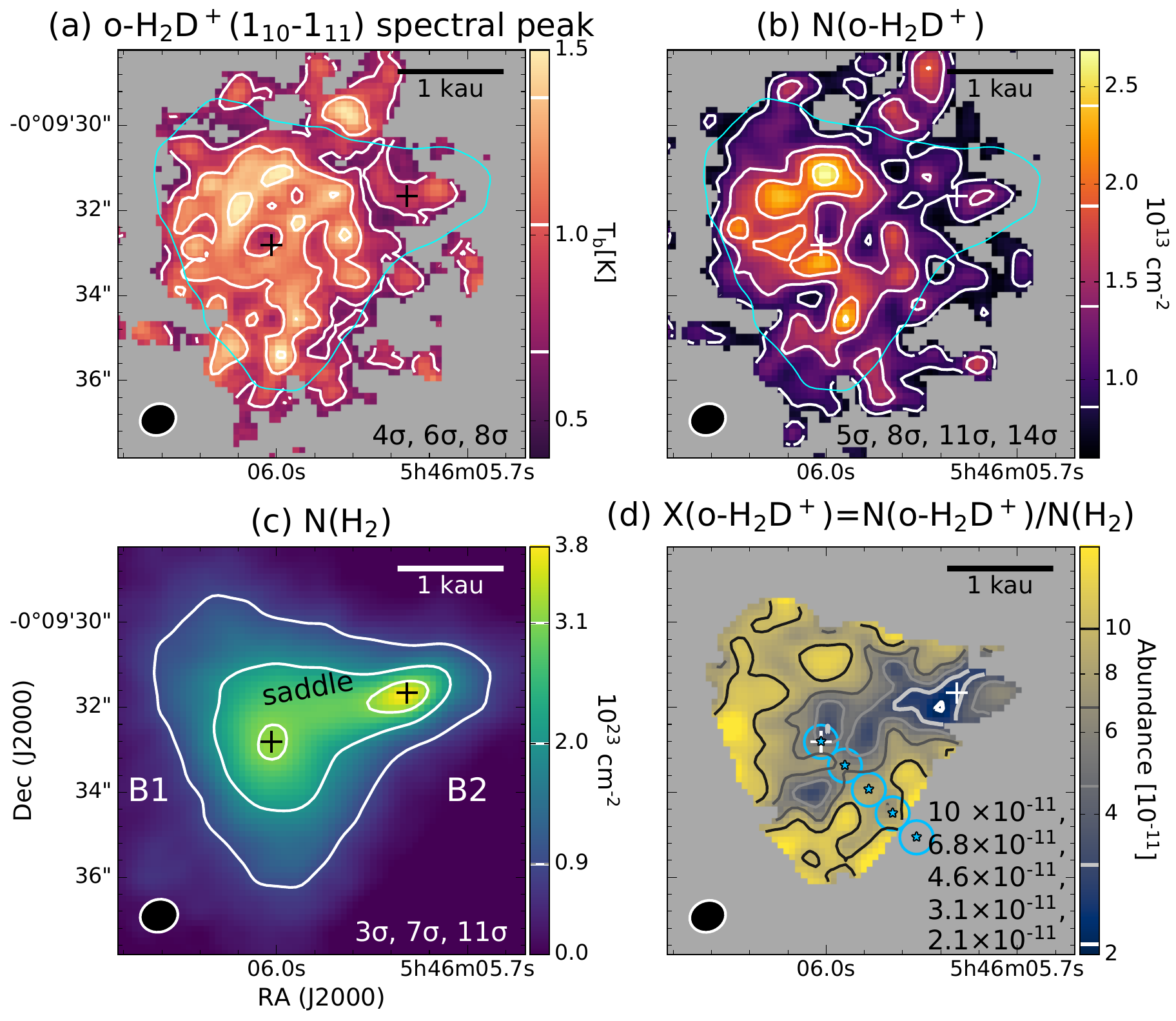}
    \caption{
    Ortho-\HtDp column density analysis assuming LTE with $T_{\rm ex}=8$~K. Calculations are limited to regions with \oHtDp(\oHtDpgrd) integrated intensities of ${\rm S/N} > 3$. 
    \textbf{(a)} Spectral peak map overlaid with contour levels at 4$\sigma$, 6$\sigma$, and 8$\sigma$, where 1$\sigma=0.17$~K.
    These $T_{\rm b}$ contour levels correspond to optical depths, $\tau_\nu$, of the order of 0.4, 0.7, and 1.0.
    \textbf{(b)} o-\HtDp column density map, $N$(o-\HtDp), overlaid with contour levels at 5$\sigma$, 8$\sigma$, 11$\sigma$, and 14$\sigma$, where $\sigma=1.7\times10^{12}$~cm$^{-2}$.
    \textbf{(c)} Molecular hydrogen column density map, $N$(\Ht), derived from dust continuum emission assuming $T_{\rm dust}=8$~K.
    Contour levels are at 3$\sigma$, 7$\sigma$, and 11$\sigma$, 
    corresponding to $8.5\times10^{22}$, $2.0\times10^{23}$, and $3.1\times10^{23}$~cm$^{-2}$, with $\sigma=2.8\times10^{22}$~cm$^{-2}$.
    \textbf{(d)} Column abundance map, $X(\mathrm{o\text{-}H_2D^+})=N(\mathrm{o\text{-}H_2D^+})/N(\mathrm{H_2})$, masked at the cyan $N$(\Ht) 3$\sigma$ contour, 
    with logarithmically spaced $X$(o-\HtDp) contour levels annotated on the panel.
    Crosses mark the substructures B1 and B2, and cyan asterisks/circles indicate positions used for non-LTE radiative transfer modeling (Section~\ref{sec:ana_nlte}). 
    Beam sizes and spatial scales are shown in the bottom left, and top right corners, respectively.
    }
    \label{fig:LTE}
\end{figure*}

\section{Evidence of Dense Gas Tracer Depletion} \label{sec:res}

\autoref{fig:cont_mole}(a) shows the JCMT 850~\micron continuum contours, which reveal the dust distribution 
including our targeted prestellar core \souname,
and the nearby protostellar cores G205.46$-$14.56\,M2 and G205.46$-$14.56\,M1 in the Orion B molecular cloud complex \citep{Dutta20}. 
Both M2 and M1 are associated with 8~\micron emission from embedded protostars.
In contrast, \souname exhibits an emission dip at 8~\micron, which appears as an absorption feature against the bright PAH background from the diffuse medium.
This 8~\micron absorption arises where the mid-infrared extinction from the prestellar core dominates over the contribution from dust-scattered light \citep{Lefevre16},
further supporting the prestellar nature of \souname.

\autoref{fig:cont_mole}(b)--(f) show the dust continuum and integrated molecular line intensity maps of \souname.
We assess the depletion of dense gas tracers in \souname by comparing their spatial distributions with the dust continuum observed by ALMA.
Our 820~\micron continuum map in \autoref{fig:cont_mole}(b) clearly resolves two substructures, B1 and B2, with peak intensities of $5.7\pm0.5$\mJybm and $6.8\pm0.5$\mJybm, respectively.
Their brightness temperatures ($\sim$0.08~K for B1 and $\sim$0.09~K for B2) are much lower than the dust temperatures \citep[7--10~K; ][]{Sahu23}, 
indicating optically thin dust and negligible dust absorption for our line intensity maps.
Opposite to the continuum, peaking toward B1 and B2, the molecular line emissions from o-\HtDp and \NtDp display distinct morphologies.

Our o-\HtDp(\oHtDpgrd) map in \autoref{fig:cont_mole}(c) shows an annulus/arc-like structure around B1, displaying a central depletion with an emission contrast up to $\sim$6$\sigma$
(from the $\sim$5$\sigma$ minimum to the $\sim$9$\sigma$--11$\sigma$ ridge).
If we conservatively take the 7$\sigma$ contour as the full width at half depth (FWHD) of the emission hole, 
the resulting diameter is slightly larger than one synthesized beam of $\sim$0\farcs8, so the B1 depletion zone is marginally resolved.
Although the central o-\HtDp intensity minimum does not exactly coincide with the continuum peak of B1, the offset lies within the beam and likely reflects a slight deviation from spherical symmetry.
In contrast, our current resolution may be insufficient to resolve a similar annulus/arc-like o-\HtDp intensity structure toward the smaller B2, even though the continuum peak of B2 is located between two nearby o-\HtDp emission peaks.

The \NtDp(3--2) emission shows a flattened intensity profile toward B1.
The two highest-level contours (19$\sigma$ and 23$\sigma$) in \autoref{fig:cont_mole}(d) are widely separated relative to the lower-level contours, suggesting a plateau-like inner emission structure rather than a sharply peaked core.
This is indicative of central depletion partly masked by outer undepleted layers, similar to the flattened \NHth morphology observed in Oph‑H‑MM1 \citep{Pineda22}.
Given the \NtDp formation pathway, 
\begin{equation}
\ce{H2D+ + N2 -> N2D+ + H2}, \label{equ:n2dp}
\end{equation}
the widespread detection of \NtDp(3--2) implies that o-\HtDp is present in the outer regions.
Although o-\HtDp emission is indeed detected beyond the 3$\sigma$ continuum contour in \autoref{fig:cont_mole}(b),
our single-pointing o-\HtDp observation is less sensitive to the extended emission at the outskirts than the uniformly sensitive \NtDp mosaic plus TP observations.
Notably, in the innermost region, 
both o-\HtDp and \NtDp exhibit signs of depletion,
but the more pronounced decline in o-\HtDp hints at the enhanced deuterium fractionation.
This difference arises because the \Hthp isotopologues can undergo multiple deuterium substitutions via reactions with HD, progressively converting \Hthp to \HtDp, \DtHp, and finally to \Dtp, whereas deuteration of \NtHp is limited to a single substitution when \Nt reacts with \HtDp (or \DtHp, \Dtp) to form \NtDp.
Consequently, the further deuteration of the \Hthp isotopologue leads to a more significant reduction in o-\HtDp abundance at the core center.

Moreover, \NtHp emission shown in \autoref{fig:cont_mole}(e) and (f) appears to be influenced by different excitation conditions, with its low $J$=1--0 transition often being optically thick.
Brighter \NtHp(1--0) and (4--3) emission on the outer eastern side of \souname may result from outflows driven by the multiple protostellar system SSV\,63 \citep{Reipurth23}, which is marked by cyan circles in \autoref{fig:cont_mole}(a) and located about 0.07~pc southeast of \souname. 
These outflows could raise the excitation temperature and/or enhance the \NtHp abundance on that side \citep{Lis16, Pagani24, LopezVazquez25}.
This may also account for the sharper eastern edges seen in the 820~\micron continuum, as well as in the o-\HtDp and \NtDp maps.

\section{Analysis} \label{sec:ana}

\subsection{Column Density Calculation} \label{sec:ana_lte}

\begin{figure*}[th]
    \centering
    \includegraphics[width=0.9\textwidth]{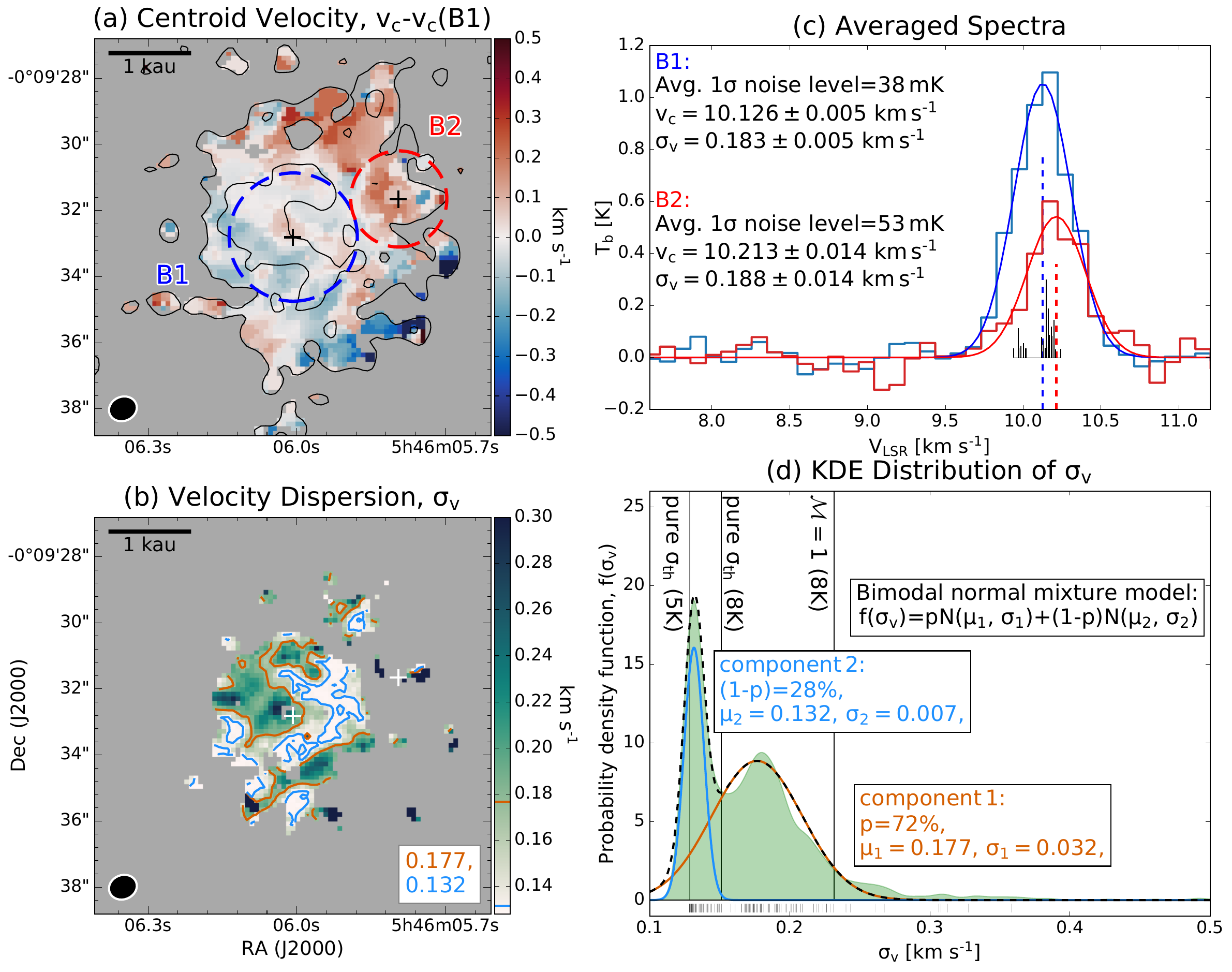}
    \caption{\soufullname kinematics traced by \oHtDp. 
    \textbf{(a)} Centroid velocity map, using data with 
    $\geq$3$\sigma$ detection on the integrated-intensity map (with the 3$\sigma$ and 7$\sigma$ contours overlaid; see \autoref{fig:cont_mole}(b)).
    \textbf{(b)} Velocity dispersion map, only using data with $\geq$5$\sigma$ detection at the spectral peaks.
    \textbf{(c)} Averaged spectra for B1 (blue) and B2 (red), extracted from the regions enclosed 
    by the corresponding dashed circles in panel (a), with fitted Gaussian profiles and annotated parameters. 
    The centroid velocity of B1 serves as the velocity reference in panel (a). The normalized o-\HtDp(\oHtDpgrd) HFS pattern is also shown in black beneath the B1 spectrum.
    \textbf{(d)} KDE distribution of velocity dispersions with a rug plot. The fitted bimodal model and its parameters are annotated, with the two dispersion peaks overlaid on panel (b) as contours.
    Vertical lines show the expected velocity dispersion including channel broadening for pure thermal cases and $\mathcal{M}=1$ cases assuming 5~K and/or 8~K (see \autoref{app:dispersion}).
    }
    \label{fig:mom}
\end{figure*}

\begin{figure*}[tbh]
    \centering
    \includegraphics[width=\textwidth]{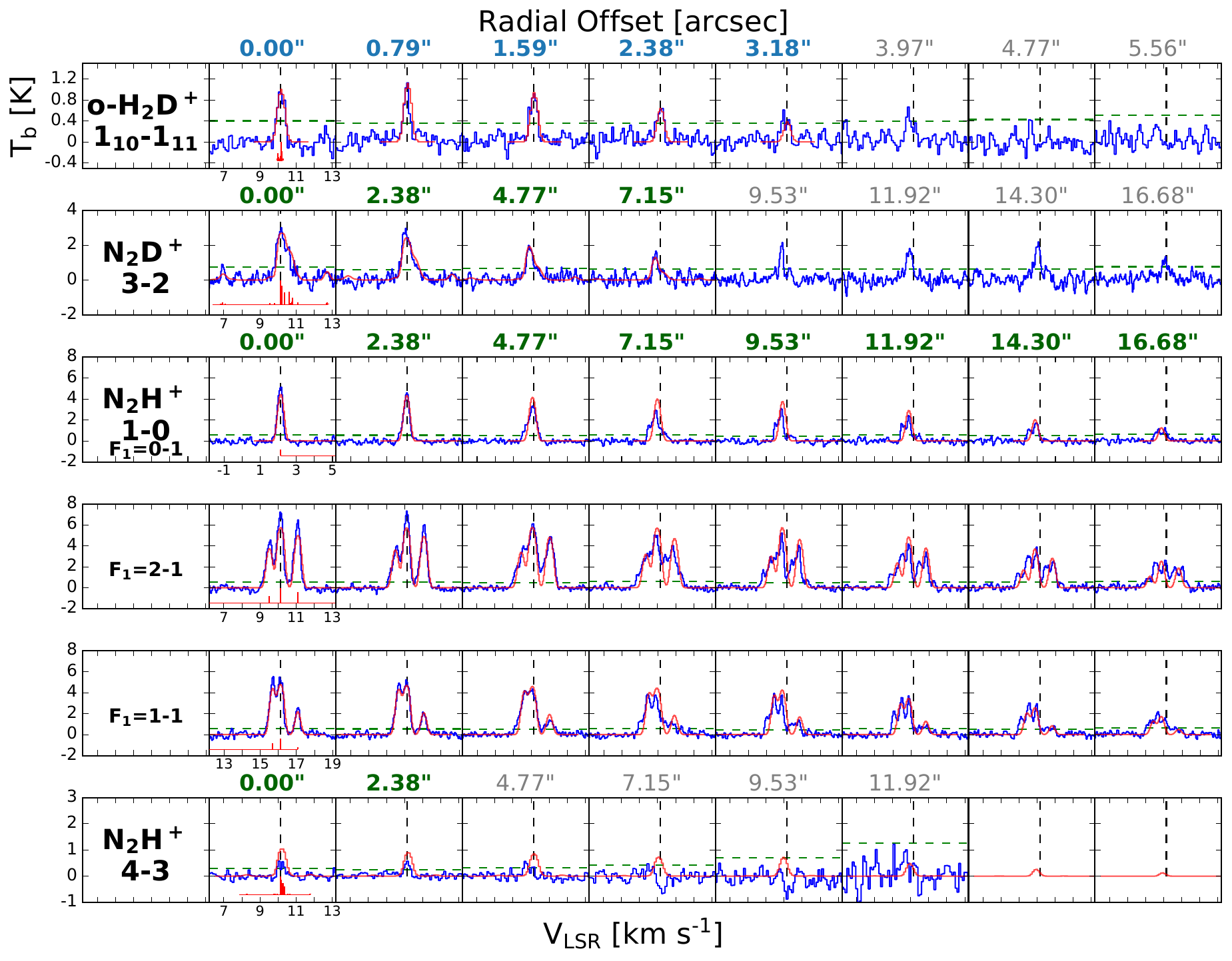}
    \caption{
    Spectral observations along the southwest cut compared with the best-fit radiative transfer model. 
    Blue spectra represent observational data, while red spectra show the models. 
    Green dashed lines mark the 3$\sigma$ noise level.
    Each row corresponds to a molecular line, with the \NtHp(1--0) line split into three rows for $F_1$-hyperfine groups. 
    Columns represent different offsets from the center of \souname B1, marked by color-coded asterisks in \autoref{fig:cont_mole}. 
    Spectra at gray offsets are excluded from modeling due to nondetections or deviations from spherical geometry. 
    Normalized HFS patterns are shown in red beneath each central spectrum. 
    Black dashed lines indicate the systemic velocity of 10.151~\kms. 
    For \NtHp(1--0), additional black dashed lines highlight the $F_1,F$=(0,1--1,2) and (1,2--1,2) components.
    }
    \label{fig:spectra}
\end{figure*}

We first determine the o-\HtDp column density, $N$(o-\HtDp), under the local thermal equilibrium (LTE) condition and the \Ht column density, $N$(\Ht), assuming optically thin 820~\micron continuum emission (Section \ref{sec:res}). 
The o-\HtDp(\oHtDpgrd) line is thermalized given its critical density of $1.1\times10^5$~cm$^{-3}$ at 8--10~K \citep{Hugo09}, well below the averaged density within the central flat region of \souname \citep[$7\times10^6$~cm$^{-3}$;][]{Sahu23}.
The detailed derivations are provided in \autoref{app:colden}. 

\autoref{fig:LTE}(a) shows the brightness temperature ($T_{\rm b}$) at the o-\HtDp(\oHtDpgrd) spectral peak.
For simplicity, we assume a uniform excitation temperature ($T_{\rm ex}$) across the core.
Given the maximum $T_{\rm b}$ of $1.52\pm0.17$~K at the northern ridge,
the lower limit for $T_{\rm ex}$ for a non-maser line is 7.1~K to find a real (non-complex) value for the optical depth ($\tau_\nu$; see \autoref{equ:tau}).
Accordingly, we adopt $T_{\rm ex}=8$~K, 
and find that the optical depth remains moderately thin ($\tau_\nu\lesssim1$) throughout \souname.
\autoref{fig:LTE}(b) presents the $N$(o-\HtDp) map. 
\autoref{fig:LTE}(c) shows the $N$(\Ht) map adopting a dust temperature ($T_{\rm d}$) of also 8~K,
considering efficient dust--gas coupling \citep{Goldsmith01}.
Then \autoref{fig:LTE}(d) shows the o-\HtDp column abundance map, $X(\mathrm{o\text{-}H_2D^+})=N(\mathrm{o\text{-}H_2D^+})/N(\mathrm{H_2})$.

The o-\HtDp column abundance is about $10^{-10}$ at the outskirts, with several local maxima around B1.
Toward the inner regions, including the continuum saddle, the abundance decreases and exhibits two distinct minima located near the continuum peaks of B1 and B2, each within the $\sim$0\farcs8 beam size.
The beam-averaged abundances are $X(\mathrm{o\text{-}H_2D^+})=(5.0\pm0.7)\times10^{-11}$ toward B1 and $(3.0\pm0.6)\times10^{-11}$ toward B2,
corresponding to column depletion factors of $\sim$2 for B1 and $\sim$3 for B2 relative to the outskirts value of  $X(\mathrm{o\text{-}H_2D^+})\sim10^{-10}$.

\subsection{Gas Kinematics}\label{sec:ana_gas}

We investigate the centroid velocity ($v_{\rm c}$) and velocity dispersion ($\sigma_v$) from o-\HtDp in \souname.
We only consider positions with a $\geq$3$\sigma$ detection in the integrated intensity for $v_{\rm c}$ measurements, and require a $\geq$5$\sigma$ detection at the spectral peak for a reliable $\sigma_v$ determination.
Toward the B2 center, the
weaker o-\HtDp signal prevents a reliable determination of $\sigma_v$.

\autoref{fig:mom}(a) shows the centroid velocity map
relative to B1 at 10.126~\kms, and
\autoref{fig:mom}(b) reveals that the velocity dispersion remains mostly below 0.3~\kms.
Due to o-\HtDp depletion, we average spectra including the undepleted regions (dashed circles in \autoref{fig:mom}(a)) as proxies for kinematic measurements for B1 and B2,
with the results shown in \autoref{fig:mom}(c).
We fit the spectra with a single Gaussian because the hyperfine structure (HFS) of o-\HtDp(\oHtDpgrd) is not spectrally resolved.
For reference, we adopt the component frequency offsets and normalized strengths tabulated by \citet{Jensen97} to plot the HFS pattern in \autoref{fig:mom}(c).
The fits indicate that B1 and B2 have similar centroid velocities (within a channel width $\delta v_{\rm ch}$ of 0.098~\kms) and consistent velocity dispersions (within their fitting uncertainties).
However, one can observe larger core-scale velocity gradients. Redshifted emission is seen on the northwest side and blueshifted emission on the southeast side, indicating a rotation that deviates slightly from the B1--B2 direction. A higher-sensitivity observation with a larger spatial coverage is required to investigate this core rotation feature further.

\autoref{fig:mom}(d) shows the kernel density estimate (KDE) of the $\sigma_v$ distribution alongside a rug plot.
Interestingly, the $\sigma_v$ sample is well described by a two-component normal mixture distribution, with components at $\sigma_v = 0.177\pm0.032$~\kms,
and $0.132\pm0.007$~\kms, 
with a mixing fraction $p=0.72$ for the higher-$\sigma_v$ component.
The higher-$\sigma_v$ component agrees with the dispersion of the averaged B1 spectrum in \autoref{fig:mom}(c) and is likely associated with the substructures. 
The measured velocity dispersion $\sigma_v$ includes broadening from thermal motions ($\sigma_{\rm th}$), 
nonthermal motions ($\sigma_{\rm nt}$), 
unresolved hyperfine splitting ($\sigma_{\rm hfs}$),
and the spectral channel response ($\sigma_{\rm ch}$); see \autoref{app:dispersion}.
Given $\sigma_{\rm hfs}=0.079$~\kms,
the HFS contribution is only $\sim$10\% of $\sigma_v$ for the higher-$\sigma_v$ component but up to $\sim$20\% for the lower-$\sigma_v$ component.
Three vertical reference lines mark the purely thermal cases at 5~K and 8~K, and $\mathcal{M}=1$ case at 8~K, computed including unresolved hyperfine structure and instrumental channel broadening.
These comparisons indicate that most nonthermal motions are subsonic at 8~K.
Notably, the lower-$\sigma_v$ component is almost purely thermal, hinting at an even lower temperature of 5--8~K in the core interior.

Although these two $\sigma_v$ components do not have clear spatial correspondences in \autoref{fig:mom}(b), the region between B1 and B2, characterized by nearly constant thermal motions, contributes to the lower-$\sigma_v$ component.
This suggests that the nonthermal motions gradually dissipate in the core interior, while the higher-$\sigma_v$ component may represent the remaining subsonic turbulence and/or infall motions toward B1.

\subsection{Non-LTE Radiative Transfer Modeling toward the Substructure B1}\label{sec:ana_nlte}

We further model the radial variations in volumetric molecular abundances of o-\HtDp, as well as \NtDp and \NtHp, toward B1, which is better spatially resolved than B2.
To minimize external influence on the eastern side, we assume the western side better reflects the core's original conditions. Accordingly, to avoid overlap with B2 to the west--northwest, we extract and analyze spectra along a southwest cut (asterisks in \autoref{fig:cont_mole}).
Spectrum pointing offsets are spaced to be comparable to the synthesized beam sizes to ensure approximate spatial independence.
For o-\HtDp, the offset is set to the circularized Band 7 beam size ($\sqrt{\theta_{\rm maj}\theta_{\rm min}}$) of 0\farcs79. 
For \NtDp(3--2), \NtHp(1--0), and \NtHp(4--3), the offset is 2\farcs38, three times the o-\HtDp offset, matching or larger than their respective beam sizes (\autoref{tab:obs}).

At each offset, we average spectra within a circular aperture of the corresponding diameter to improve S/N, as illustrated for o-\HtDp by the cyan circles in \autoref{fig:LTE}(d).
Although the central pointing is at the B1 continuum peak rather than at the $X$(o-\HtDp) minimum, the separation lies within the aperture, and the resulting difference between the aperture-averaged spectra is within the rms noise and negligible for our analysis.
For \NtDp(3--2), we use only positions whose integrated-intensity morphology appears round to avoid contamination from the north--south filament.
We therefore model the detected o-\HtDp(\oHtDpgrd), \NtDp(3--2), \NtHp(1--0), and \NtHp(4--3) emission out to maximum offsets of 3\farcs18, 
7\farcs15,
16\farcs68,
and 2\farcs38 from the center, respectively.
The extracted spectra (blue) and best-fit model profiles (red), with color-coded offsets, are shown in \autoref{fig:spectra}.
For \NtHp(1--0), two velocity components are present at $r>7$\farcs15 and merge inward; we fit the component near the B1 systemic velocity using the isolated $F_1=(0\text{--}1)$ group.

Our analysis employs a 1D spherically symmetric non-LTE Monte Carlo radiative transfer framework \citep{Bernes79}, accounting for hyperfine transition overlaps \citep{Pagani07}, with HFS-resolved collisional rate coefficients for \NtHp \citep{Lique15} and \NtDp \citep{Lin20}. 
Although the o-\HtDp(\oHtDpgrd) transition has hyperfine structure \citep{Jensen97}, the components are not spectrally resolved in our data and, to our knowledge, HFS-resolved collisional rates remain unavailable.
We therefore model o-\HtDp with a single Gaussian profile and compute its excitation using the rotational collision rates from \citet{Hugo09}.
The modeled brightness distribution is then convolved with the corresponding circular Gaussian beam before comparison with the extracted spectra.

We construct a layered spherical physical model for B1 that includes the \Ht volumetric density ($n_{\rm H_2}$), gas kinetic temperature ($T_{\rm k}$), and molecular abundances ($x_{\rm species}=n_{\rm species}/n_{\rm H_2}$) profiles, assuming a uniform nonthermal dispersion ($\sigma_{\rm nt}$). 
The model is discretized into 16 layers with two radial step sizes set by the pointing offsets: 12 inner layers with radial thicknesses 0\farcs79 (317~au for a distance of 400~pc), and 4 outer layers with thicknesses 2\farcs38 (953~au, three times thicker). 
The division at $r=9$\farcs53 (3,810~au) aligns with the extent of \NtDp emission.

The physical structure adopts the Plummer-like $n_{\rm H_2}$ profile of \souname,
\begin{equation}
n_{\rm H_2}(r)=\frac{1.1\times10^7}{1+(521~\text{au}/r)^2}~\text{cm}^{-3},
\label{equ:den}
\end{equation}
with $r$ in au. 
This profile was derived by \citet{Sahu23} from ALMA 1.3~mm data ($\sim$1\farcs2 resolution), assuming $T_{\rm d}=10$~K and accounting for interferometric filtering.
For the kinetic temperature, we start from a reference $T_{\rm k}=T_{\rm dust}=10$~K and adopt a three-step profile guided by the LTE o-\HtDp result $T_{\rm ex}=8$~K (Section~\ref{sec:ana_lte}), and the \textit{Herschel} dust temperature $T_{\rm dust}=14$~K \citep{Konyves20, Yoo23}:
\begin{equation}
T_{\rm k}(r)=
\begin{cases}
8~\text{K}, & r<1910~\text{au},\\
10~\text{K}, & 1910~\text{au}<r<3810~\text{au},\\
14~\text{K}, & 3810~\text{au}<r<7630~\text{au}.
\end{cases}\label{equ:temp_prof}
\end{equation}
The step radii are set by the observed extents of the molecular emission and represent a cooler center with warmer outer layers.
Although several sigmoid-like temperature profiles \citep[e.g.,][]{Crapsi07, Magalhaes18} are sometimes used, our angular resolution renders any monotonic profile effectively piecewise; a three-zone step function is therefore sufficient for the present data.
As o-\HtDp traces the innermost regions, we take its kinematics to represent all modeled species. 
Gaussian spectral profile fitting of the averaged o-\HtDp spectra within the 3$\sigma$ contour in \autoref{fig:cont_mole}(c) yields a systemic velocity $v_{\rm c}=10.151$~\kms and a dispersion $\sigma_v=0.183$~\kms.
From this, we infer a nonthermal dispersion $\sigma_{\rm nt}=0.130$~\kms at 
$T_{\rm k}=8$~K. We adopt this as a spatially uniform value in the model, and we likewise fix $v_{\rm c}$.
Additionally, small velocity shifts of 0.3--0.4~\kms are applied to the \NtDp(3--2) and \NtHp(1--0) spectra at $r>4$\farcs77 to account for their slight redshift at large radii.

\begin{figure*}[tbh]
    \centering
    \includegraphics[width=\textwidth]{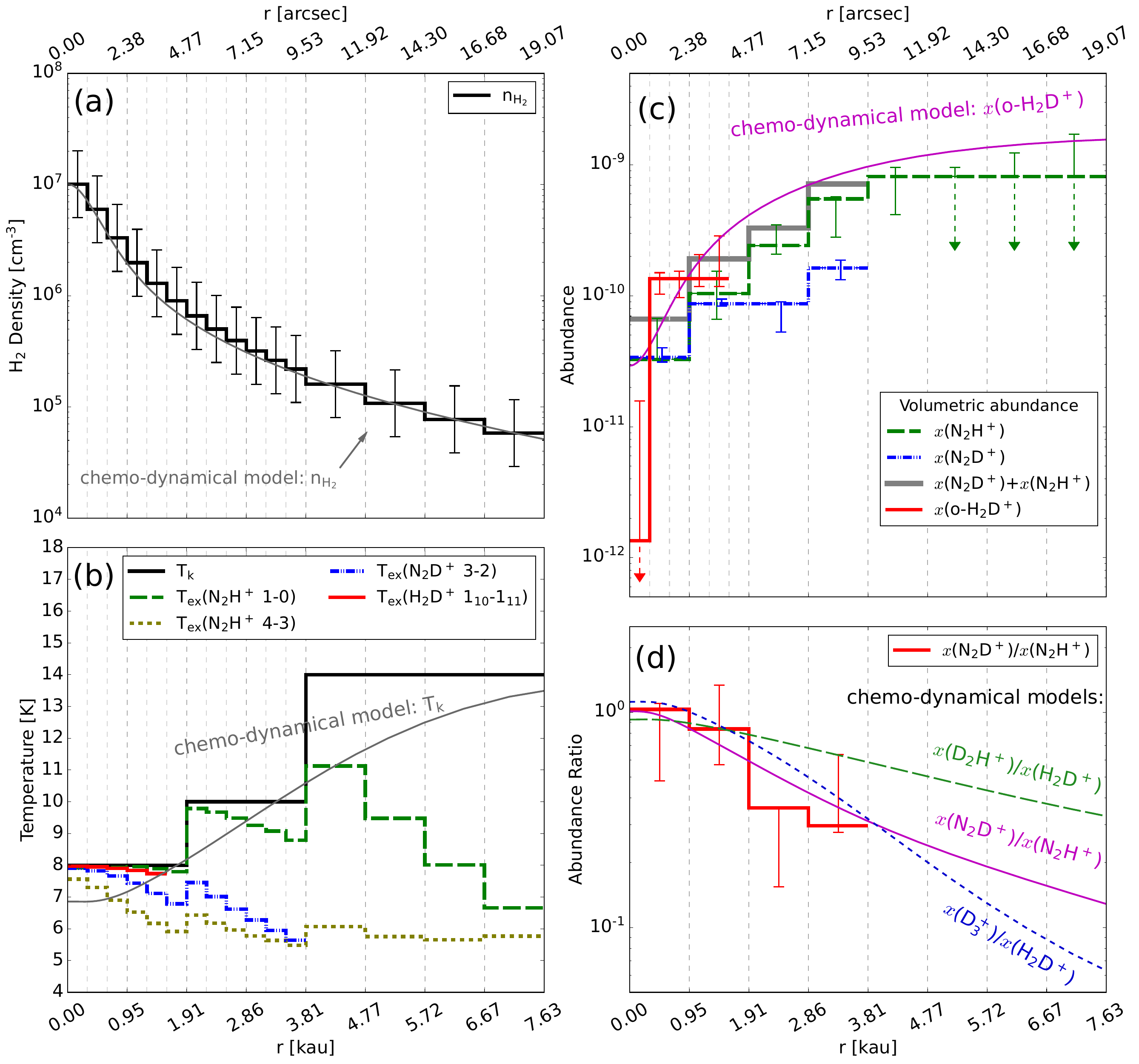}
    \caption{
    Spherical physical model of \soufullname B1 based on spectra extracted along the southwest cut in the non-LTE radiative transfer analysis. 
    The finer model grid (light gray vertical lines) resolves the $x$(o-\HtDp) profile with a step size of 0\farcs79, while the coarser model grid (dark gray vertical lines) resolves the $x$(\NtDp) and $x$(\NtHp) profiles with a step size of 2\farcs38.
    \textbf{(a)} Plummer-like $n_{\rm H_2}$ profile is adopted from \citet{Sahu23}, with error bars reflecting a factor-of-2 variation due to typical dust opacity uncertainties.
    \textbf{(b)} $T_{\rm k}$ profile is assumed to be 10~K, except for $r<1$\farcs91, where 8~K is inferred from multitransition \NtHp observations.
    The $T_{\rm ex}$ profiles derived from non-LTE modeling are shown for each spectral line.
    \textbf{(c)} Volumetric abundance profiles, including \textbf{(d)} \NtHp deuteration, are fitted to the extracted spectra (\autoref{fig:spectra}) with 1$\sigma$ error bars shown.
    Chemodynamical model results at a core age of 0.42~Ma for the $n_{\rm H_2}$, $T_{\rm k}$, $x$(o-\HtDp), and several deuteration profiles are overlaid for comparison.
    }
    \label{fig:nonLTE_chem}
\end{figure*}

Molecular abundance profiles are fitted as free parameters on a layered grid: o-\HtDp is resolved more finely (0\farcs79 steps), whereas \NtHp and \NtDp use 2\farcs38 steps, reflecting their effective resolutions.
Abundances are optimized sequentially from the outermost to the innermost layer by minimizing the chi-squared statistic,
\begin{equation}
	\chi^2(\text{molecule})=\sum_{i_{\rm J\text{--}J'}} \sum_{j_{\rm ptg}} \sum_{k_{\rm ch}} \left(\frac{I_{{\rm obs},k,j,i}-I_{{\rm model},k,j,i}}{\sigma_{{\rm ch},j,i}}\right)^2,
\end{equation}
where the sums run over spectral channels ($k$), pointing offsets ($j$), and rotational transitions ($i$; for \NtHp this includes $J$=1--0 and $J$=4--3). 
Here $I_{{\rm obs},k,j,i}$ and $I_{{\rm model},k,j,i}$ are the observed and modeled intensities, and $\sigma_{{\rm ch},j,i}$ is the per-channel rms noise.
The resulting best-fit parameters from our layered model are presented as step functions in \autoref{fig:nonLTE_chem}. We note that, for comparison, we overplot predictions from the chemodynamical analysis presented in Section~\ref{sec:ana_chemdyn} as continuous curves computed on a much finer Lagrangian grid.

\autoref{fig:nonLTE_chem}(a) shows the adopted Plummer density profile, whereas \autoref{fig:nonLTE_chem}(b) shows the three-step kinetic temperature profile and the excitation temperature ($T_{\rm ex}$) is also shown for each line (for \NtHp and \NtDp, this value corresponds to its strongest hyperfine component).
While the similarity between $T_{\rm ex}(\mathrm{o\text{-}H_2D^+\text{\,} 1_{10}\text{--}1_{11}})$ and $T_{\rm k}$ validates our previous assumption about LTE in Section \ref{sec:ana_lte}, the departure of $T_{\rm ex}$ for the \NtHp isotopologue lines from $T_{\rm k}$ highlights their
subthermal excitation and the importance of non-LTE treatment.

The best-fit abundances and the \NtHp deuteration ($x$(\NtDp)/$x$(\NtHp)) are shown with 1$\sigma$ error bars on \autoref{fig:nonLTE_chem}(c) and (d).
Quoted uncertainties are 1$\sigma$ defined by $\Delta\chi^2=1$ for the abundance in each layer.
Our results indicate a central o-\HtDp depletion zone with a diameter of 1\farcs58, 
$\sim$600~au at 400~pc.
Since the o-\HtDp emission hole 
is marginally resolved (see Section~\ref{sec:res}), this $\sim$600~au diameter should be regarded as an upper limit to the depletion size.

Radiative transfer modeling shows that the o-\HtDp emission is dominated by the outer layers ($316\text{\,}\mathrm{au}<r<1,590\text{\,}\mathrm{au}$) with $x(\mathrm{o\text{-}H_2D^+})=1.35\times10^{-10}$.
To reproduce the observed central spectrum, the model requires $x(\mathrm{o\text{-}H_2D^+})\leq1.35\times10^{-12}$ in the innermost layer;
further decreases have little effect due to the outer layers' emission contribution along the central sightline.
We therefore adopt $1.35\times10^{-12}$ as the fiducial central abundance for \autoref{fig:nonLTE_chem}. 
The lower side of $x(\mathrm{o\text{-}H_2D^+})$ at the central layer is instead unconstrained,
so we report only a one-sided upper limit of $1.57\times10^{-11}$, 
which would produce a central spectrum 1$\sigma$ stronger than observed.
Using this 1$\sigma$ upper limit relative to the outer-layer abundance, we conservatively derive a lower limit on the volumetric depletion factor of $\sim$9.

Surprisingly, our analysis reveals a higher central $x$(\NtDp)/$x$(\NtHp) ratio of 
$1.03_{-0.56}^{+0.07}$,
compared to $\sim$0.1--0.7 observed by low-resolution single-dish studies on other starless/prestellar cores \citep[e.g.,][]{Crapsi05, Pagani07}.
This indicates that the o-\HtDp depletion is driven by further deuterium fractionation in the core center. 
However, while the $x$(\NtDp)/$x$(\NtHp) ratio increases, the total \NtHp isotopologue abundance ($x$(\NtDp) + $x$(\NtHp)) decreases toward the center
over a larger region (with a diameter of $\sim$8,000~au), which can be explained by the widespread \Nt freeze‐out in the core interior.

\subsection{Chemodynamical Modeling}\label{sec:ana_chemdyn}

To interpret our observations of significant depletion of o-\HtDp and enhanced \NtDp/\NtHp fractionation, we simulate its chemistry using a chemodynamical model adopted from \citet{Pagani13}.
The model couples the time-dependent gas-phase deuterium chemistry with the hydrodynamical evolution during core contraction, solved on a 1D Lagrangian (mass-coordinate) grid that follows concentric shells \citep{Lesaffre05}.
The chemical network is specifically tailored for prestellar cores, incorporating all spin states of the \Ht and \Hthp isotopologues and assuming ``complete depletion" for heavy species (except for CO and \Nt).

\citet{Pagani13} presented two fiducial models with different contraction timescales: a fast model (comparable to the freefall time), and a slow model (about 10 times slower, mimicking ambipolar-diffusion timescales).
Both simulate core contraction from an initially uniform-density sphere ($n_{\rm H_2}=5\times10^3$~cm$^{-3}$) at an initial temperature of 10~K. 
Within each model, variations in parameters like the initial o-\Ht/p-\Ht value (OPR$_0$(\Ht)), the cosmic-ray ionization rate ($\zeta$), and the grain size ($a_{\rm d}$) were also tested.
Notably, the resulting $x$(\NtDp)/$x$(\NtHp) profiles differ between the two models.
In the slow-contraction model, deuterium fractionation reaches full development, while in the fast case, it builds gradually.

Our observed $x$(\NtDp)/$x$(\NtHp) profile in \souname B1 favors the fast-contraction model since the slow-contraction model predicts larger $x$(\NtDp)/$x$(\NtHp) values.
We therefore simulate B1 by adopting the fast model parameters \citep[see][Sections 5.1 and 5.2]{Pagani13}, and standard values of OPR$_0(\text{H$_2$})=3$, $\zeta=3\times10^{-17}$~s$^{-1}$, and $a_{\rm d}=0.1$~\micron.
The simulation is stopped when the central density reaches $n_{\rm H_2}=1.1\times10^7$~cm$^{-3}$ for B1.
To match the Plummer-like density profile (\autoref{equ:den}),
we set an initial sphere radius of 0.15~pc.
The model assumes constant CO and \Nt abundances (i.e., they are not fully depleted) that are chemically linked to those of \NtHp, \NtDp, and o-\HtDp. We then adjust these to $x(\mathrm{CO})=7\times10^{-6}$ and $x(\mathrm{N_2})=8\times10^{-7}$ so that the resulting $x$(\NtDp)/$x$(\NtHp) and $x$(o-\HtDp) profiles match our observations.

Our simulations with core ages in the range 0.418--0.425~Ma can reproduce the consistent results with the observed profiles, indicating a representative core age of $\sim$0.420~Ma.
The resulting radial profiles of $n_{\rm H_2}$, $T_{\rm k}$, $x$(o-\HtDp), and selected deuteration ratios, evaluated at 0.422~Ma when the central density equals $1.1\times10^7$~cm$^{-3}$, are overplotted in \autoref{fig:nonLTE_chem} as continuous curves for direct comparison with the stepwise profiles from the layered non-LTE radiative transfer analysis. 
A quantitative comparison and the implications for further deuteration within the \Hthp isotopologues are discussed in Section~\ref{sec:dis_h3p}.

\section{Discussion} \label{sec:dis}

\subsection{Further deuteration of $H^+_3$}\label{sec:dis_h3p}

\begin{figure*}[th]
    \centering
    \includegraphics[width=0.9\textwidth]{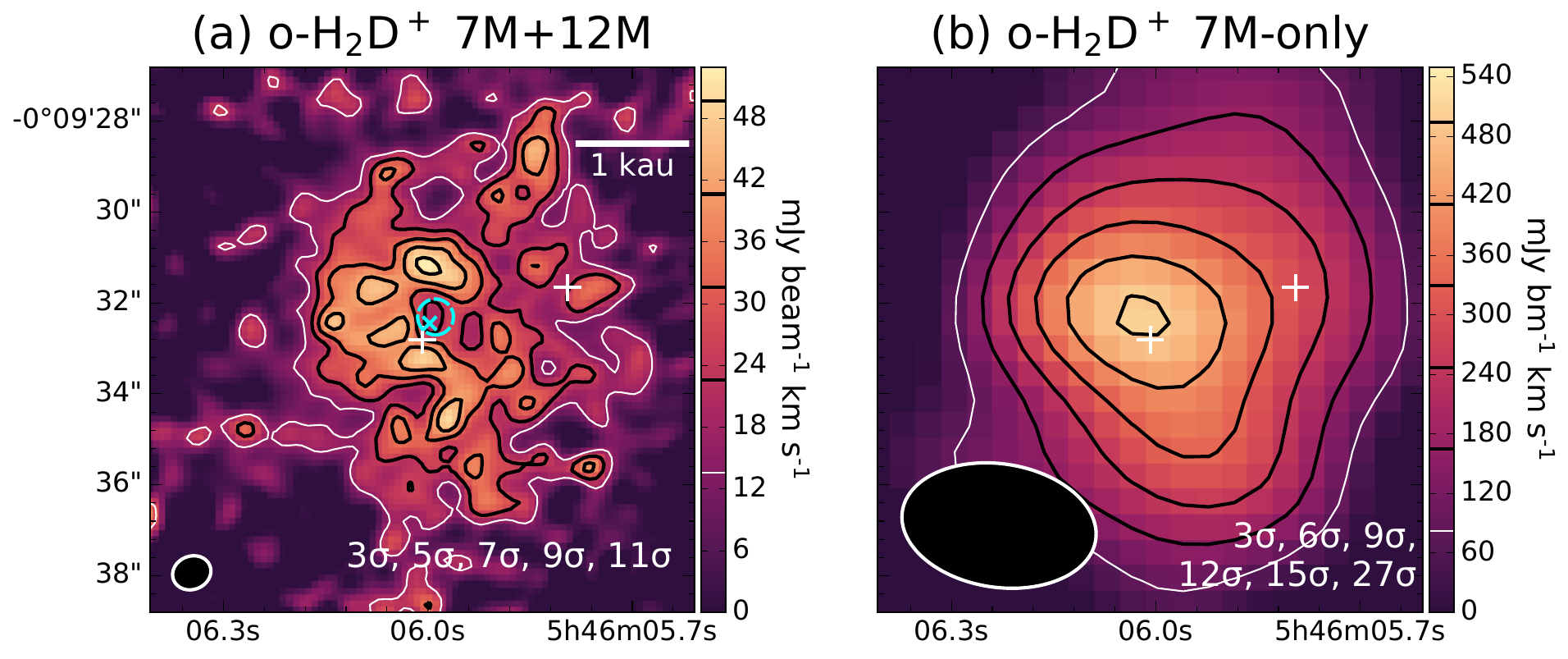}
    \caption{Comparison of the ALMA 7\,m+12\,m combined map and the 7\,m-only map of o-\HtDp(\oHtDpgrd).
    \textbf{(a)} 7\,m+12\,m combined integrated-intensity map, identical to \autoref{fig:cont_mole}(c). The dashed cyan circle marks the circularized Band~7 beam (0\farcs79) and approximates the FWHD of the depletion zone, while the cyan “$\times$” marks the intensity minimum.
    \textbf{(b)} 7\,m--only integrated-intensity map over the same velocity range as (a), with synthesized beam of $4\farcs30\times2\farcs72$ (P.A.\,$=81\fdg7$), and
    the 1$\sigma$ rms of 27~\mJybmkms. 
    To maximize contrast with the color scale, the 3$\sigma$ o-\HtDp contour is drawn in white, while higher contours are black. 
    Beam sizes, contour levels, and crosses marking B1 and B2 are indicated in each panel. 
    Color images are primary-beam-corrected, whereas contours are drawn from primary-beam-uncorrected images to preserve uniform noise.}
    \label{fig:h2dp_comp}
\end{figure*}

\autoref{fig:nonLTE_chem} shows that 
the modeled $x$(o-\HtDp) and $x$(\NtDp)/$x$(\NtHp) profiles (continuous curves) agree with our observationally derived abundances (stepwise profiles). 
The modeled $T_{\rm k}$ profile is also consistent with our assumed three-step temperature profile (\autoref{equ:temp_prof}).
These results confirm that the depletion of o-\HtDp is linked to advanced deuteration,
evidenced by the high central $x$(\NtDp)/$x$(\NtHp) ratio and by simulated abundance ratios of both \DtHp and \Dtp relative to o-\HtDp 
peaking toward the center at values of order unity. 
We note that the modeled $x$(o-\HtDp) rises outside the observed radius ($r>1,590$~au), but those outer values are likely overestimated because
the model fixes $x$(CO$)=7\times10^{-6}$ to mimic depletion, a depletion factor $\sim$14 relative to the canonical $10^{-4}$ \citep{Pineda10}. At large radii, CO is expected to be less depleted, which would increase $x$(CO) and lower $x$(o-\HtDp) given that CO is the main destroyer of \Hthp isotopologues.

However, the modeled $x$(o-\HtDp) is only marginally consistent with the observed abundance, remaining just within the 
2$\sigma$ upper limit in the $\sim$600-au depletion zone (the innermost o-\HtDp layer in \autoref{fig:nonLTE_chem}(c)).
This discrepancy is similar to recent o-\HtDp (\oHtDpgrd) results toward IRAS\,16293E, where an o-\HtDp depletion zone is spatially anticorrelated with p-\DtHp (\pDtHpgrd) emission \citep{Pagani24}.
Given the sample of only two sources (IRAS\,16293E and \souname B1), we speculate that the discrepancy between the observations and the chemical models arises from limitations in the current chemical network, such as incomplete chemical or collisional rate coefficients, rather than from observational uncertainties \citep[see also the discussion in][]{Pagani24}.

To resolve such depletion zones, 
the FWHD of the emission hole should be at least twice the beam size 
(i.e., FWHD $\gtrsim 2\theta_{\rm res}$).
Thus, the o-\HtDp depletion zones are marginally resolved in both \souname and IRAS\,16293E.
In \souname, the FWHD is comparable to the circularized Band\,7 beam of 0\farcs79 (see Section~\ref{sec:res} and \autoref{fig:h2dp_comp}(a)), 
whereas in IRAS\,16293E the FWHD is only about half of the 14\arcsec JCMT beam \citep[see Figure\,2 from][]{Pagani24}.
This implies that the depletion-zone diameters derived via radiative transfer are upper limits 
(1\farcs58 or $\sim$600~au for \souname, and 16\farcs2 or $\sim$2,300~au for IRAS\,16293E), 
and that the reported emission contrasts 
($\sim$6$\sigma$ for \souname, and $\sim$2.3$\sigma$ for IRAS\,16293E) are sensitive to angular resolution.
If better resolved, one would expect a narrower and deeper depletion zone.

In particular, although \souname shows a stronger contrast than IRAS\,16293E, 
no depletion is seen in the 7\,m--only map (3\farcs4 beam; 1,360~au at 400~pc) as shown in \autoref{fig:h2dp_comp}(b).
This indicates that the \souname emission hole is completely hidden by beam dilution at 1,360-au resolution, which is still smaller than the JCMT linear resolution toward IRAS\,16293E.
Taken together, this suggests that IRAS\,16293E hosts a larger and more strongly depleted zone than \souname,
hence may be more evolved.
Additional p-\DtHp observations toward \souname are needed to confirm these comparisons. 
Nonetheless, the transition from o-\HtDp to more heavily deuterated species suggests that p-\DtHp(\pDtHpgrd) 
would serve as a tracer of the densest, most evolved core regions.

\subsection{Fast and Slow Core Contraction}\label{sec:fast_slow}

\begin{figure*}[ht]
    \centering
    \includegraphics[width=\textwidth]{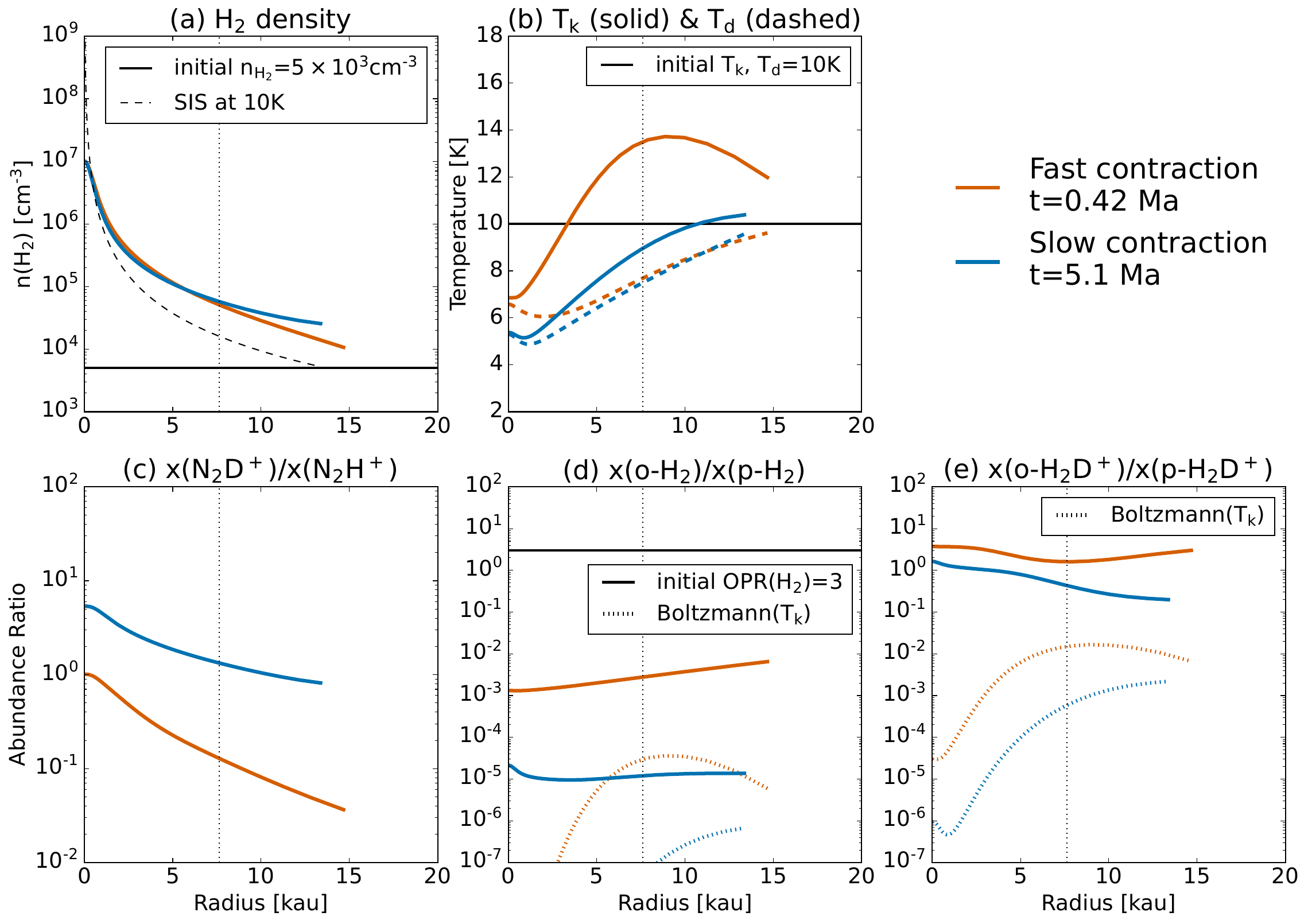}
    \caption{Comparison of fast and slow-contraction chemodynamical models.
    (a) \Ht number density, $n_{\rm H_2}$. 
    (b) Gas kinetic temperature, $T_{\rm k}$ (solid), and dust temperature, $T_{\rm d}$ (dashed).
    (c) Deuteration ratio, $x$(\NtDp)/$x$(\NtHp).
    (d) Ortho-to-para ratio of \Ht, OPR(\Ht).
    (e) Ortho-to-para ratio of \HtDp, OPR(\HtDp).
    Orange and blue curves correspond to snapshots at $t=0.42$~Ma (fast) and $t=5.1$~Ma (slow), respectively, chosen when both reach a central density of $n_{\rm H_2}(0)=1.1\times10^7$ cm$^{-3}$.
    The initial uniform density $5\times10^3$ cm$^{-3}$, a 10~K singular isothermal sphere (SIS) reference, the initial $T_{\rm k}=T_{\rm d}=10$~K, and the initial OPR(\Ht) = 3 are indicated.
    Boltzmann OPR(\Ht) and OPR(\HtDp) corresponding to the $T_{\rm k}$ profile are noted.
    The dotted vertical line marks the outer boundary used in the non-LTE radiative-transfer modeling (7.63~kau).}
    \label{fig:fastNslow}
\end{figure*}

\autoref{fig:fastNslow} compares a slow-contraction chemodynamical model with the fast-contraction model shown in \autoref{fig:nonLTE_chem}. 
The slow case reaches the same central density of $1.1\times10^7$~cm$^{-3}$ (see \autoref{fig:fastNslow}(a)) at an age of 5.1~Ma, about 10 times longer than the fast case, which attains it at 0.42~Ma. 
The fast case age is comparable to the freefall time of 0.43~Ma at an initial density of $5\times10^3$~cm$^{-3}$. 
The slow case is intended to mimic an ambipolar-diffusion timescale, often taken to be roughly 10 times the freefall time \citep[e.g.,][]{Das21}.

Because the slow model evolves for a longer time before reaching the same $n_{\rm H_2}$, gas--grain collisional coupling cools the gas closer to the dust temperature, bringing $T_{\rm k}(r)$ nearer to $T_{\rm d}(r)$, as shown in \autoref{fig:fastNslow}(b).
Under these conditions of low temperatures and a longer contraction time, the deuterium fractionation is allowed to progress further. 
\autoref{fig:fastNslow}(c) shows that the deuteration of \NtHp grows by factors of about 5--10 across radii. 
This is reflected by a stronger reduction of the \Ht ortho-to-para ratio in \autoref{fig:fastNslow}(d), which can drop by up to 2 orders of magnitude in the slow case. A lower OPR(\Ht) favors further deuteration of the \Hthp isotopologues and, consequently, higher \NtDp/\NtHp ratios \citep{Pineau91, Pagani92chem, Pagani09b, Pagani13, Bovino21}.

In both fast and slow models, OPR(\Ht) and OPR(\HtDp) remain far from their Boltzmann thermalized values \citep{Flower04b, Hugo09},
\begin{align}
    \frac{x(\mathrm{o\text{-}H_2})}{x(\mathrm{p\text{-}H_2})}=&9\exp\!\left(-\frac{170.5}{T_{\rm k}}\right),\\
    \frac{x(\mathrm{o\text{-}H_2D^+})}{x(\mathrm{p\text{-}H_2D^+})}=&9\exp\!\left(-\frac{86.4}{T_{\rm k}}\right),
\end{align}
as shown in \autoref{fig:fastNslow}(d) and (e).
\Ht is thought to form on grain surfaces with the statistical OPR of 3, after which OPR(\Ht) declines via reactions with H$^+$ and \Hthp.
However, \citet{Flower06a} showed that the  thermalization is efficient only down to $\sim$20~K.
At $T_{\rm k}\lesssim20$~K, OPR(\Ht) remains out of thermalization and thus retains a memory of the core's chemical/thermal history, making deuteration a tracer of past contraction rather than of the instantaneous temperature.
Although chemistry accelerates at very high densities, prestellar cores spend most of their lifetimes at relatively low densities.
There is ample time for this ``chemical clock'' to operate (e.g., in our fast model the central density stays below $10^6$~cm$^{-3}$ until $t\simeq0.41$~Ma, whereas in the slow model it remains below $10^6$~cm$^{-3}$ until $t\simeq5.0$~Ma).
Moreover, although OPR(\Ht) can approach its thermal value in diffuse clouds via H$_2$-H$^+$ collisions \citep{Shull21}, in cold dense cores the continual \Ht formation on grains replenishes o-\Ht, while low temperatures suppress ortho-to-para conversion, preventing full thermalization.

The slow and fast cases are identical except for the relaxation time $t_{\rm relax}$ used in an additional linear drag term \citep{Lesaffre05} in the radial velocity equation for $u(r, t)$,
\begin{equation}
\frac{du}{dt} = \cdots - \frac{u}{t_{\rm relax}},
\end{equation}
which damps the infall speed (in the absence of other forces, $u(t)\propto \exp[-t/t_{\rm relax}]$). 
In the fast case, we set $t_{\rm relax}=10$~Ma, much longer than the core age of 0.42~Ma, so the drag is negligible and serves only to suppress numerical shocks, as in \citet{Pagani13}. 
In the slow case, we set $t_{\rm relax}=0.01$~Ma, which strongly brakes the collapse and mimics magnetic support by ambipolar diffusion.
Frictional heating associated with this drag is not included, so the slow-contraction temperature profile should be viewed as a lower bound, whereas the fast-contraction $T_{\rm k}$ profile provides an upper bound. 
Both models remain well below $T_{\rm k}\approx 20$~K. 
Therefore, this approximation does not affect our conclusion that deuteration can develop further in the slow case.

\subsection{Turbulent-dominated core fragmentation}

Based on the B1--B2 separation of $\sim$1,200~au, \citet{Sahu21} has suggested that these substructures formed via turbulent-dominated (gravoturbulent) core fragmentation, as predicted by \citet{Offner10, Offner12}.
The turbulence activity was later also suggested by \citet{HsuSY25}.
Our results also further support this scenario.
The core age of 0.42~Ma from our chemodynamical modeling, contracting from an initial density of $5\times10^3$~cm$^{-3}$, is comparable to the free‑fall time of 0.43~Ma at that density, indicating rapid contraction.

The core ages in such models are also dependent on the initial OPR(\Ht),
which is plausibly in the range $\sim$0.5--3 from diffuse-cloud constraints \citep{Crabtree11} and can be as low as $\sim$0.1 in some simulations of filament formation \citep{Lupi21}. 
In our case, adopting OPR$_0$(\Ht)=0.1 overproduces the central $x$(\NtDp)/$x$(\NtHp) by a factor of $\sim$3 at the epoch when the central density matches that of \souname, disfavoring this initial value for this core.

Moreover, the gas between B1 and B2 exhibits an almost purely thermal velocity dispersion (Section \ref{sec:ana_gas}), 
implying that turbulence has dissipated rapidly, within no more than a few free‑fall times \citep[e.g.,][]{Nakano98, MacLow99, MacLow04}.
These rapid dynamical timescales provide supporting evidence for turbulent fragmentation, as opposed to the slower, quasi‑static core contraction characterized by ambipolar diffusion \citep[$\sim10\times$ the freefall time; e.g.,][]{Das21}.

Lastly, we find that both substructures are subvirial (i.e., $\alpha<1$) with virial parameters of $\alpha=0.40\pm0.05$ for B1 and $\alpha=0.70\pm0.17$ for B2 (see \autoref{app:mass} and \autoref{tab:masses}), indicating they are gravitationally bound. 
In the absence of additional support such as magnetic fields, these substructures are expected to collapse.
However, whether the system will form a wide protobinary at the current separation remains open.
Subsequent inspiral could also shrink the orbit and produce a close protobinary as shown by \citet{Kuruwita23}.

\section{Conclusions and Summary} \label{sec:con}

We present the first interferometric map of o-\HtDp depletion in the prestellar core \soufullname using ALMA. 
Our high-sensitivity 820~\micron dust continuum and molecular line observations reveal two distinct substructures, B1 and B2, with evidence for a $\sim$600~au o-\HtDp depletion zone toward B1. 
The main conclusions are the following.
\begin{enumerate}

\item Significant o-\HtDp depletion is found toward \souname B1.
The o-\HtDp depletion shows a $\sim$6$\sigma$ contrast between the minimum and the surrounding ridge, but the size is only marginally resolved, with the FWHD comparable to the Band~7 synthesized beam.
We therefore inferred an upper limit to the depletion-zone diameter of $\sim$600~au.
Monte Carlo radiative transfer requires a strongly reduced central abundance (fiducial $x(\mathrm{o\text{-}H_2D^+})=1.35\times10^{-12}$ with a one-sided $1\sigma$ upper limit of $1.57\times10^{-11}$), whereas the outer layers have $x(\mathrm{o\text{-}H_2D^+})=1.35\times10^{-10}$.
This implies a volumetric depletion factor $\gtrsim9$ (and $\sim$2 in column).

\item o-\HtDp depletion is linked to advanced deuteration.
The abundance profiles of o-\HtDp, \NtHp, and \NtDp in the inner core are derived with a discrete spherical-shell model using Monte Carlo radiative transfer and are broadly consistent with the self-consistent chemodynamical model. 
The high central ratio $x(\mathrm{N_2D^+})/x(\mathrm{N_2H^+})=1.03^{+0.07}_{-0.56}$ indicates the advanced deuteration of \Hthp.
However, the main exception is that the chemodynamical model only marginally reproduces the observed central o-\HtDp depletion with the observational 2$\sigma$ limit. 
This suggests that additional deuteration may be at play and that p‑\DtHp observations are needed for further constraints.

\item Kinematics are largely subsonic with a quiescent intersubstructure region.
The o-\HtDp velocity dispersions are mostly subsonic at $T_{\rm k}\sim8$~K across the core.
The $\sigma_v$ distribution is well described by a two-component mixture: a dominant, broader component likely reflects residual turbulence and/or infall toward B1, and a secondary, near-thermal component (implying $T_{\rm k}\sim5$~K) arising between B1 and B2.
Contributions from unresolved hyperfine splitting and the channel response are included in the interpretation.

\item Rapid, turbulence-dominated fragmentation is favored.
Chemodynamical modeling yields a representative core age of $\sim0.42$~Ma, comparable to the freefall time at the initial density. 
Together with the nearly thermal linewidths between B1 and B2, this suggests rapid gravoturbulent fragmentation rather than slow, quasistatic contraction, with turbulence likely dissipating in no more than a few freefall times.
Both substructures are subvirial ($\alpha=0.40\pm0.05$ for B1; $\alpha=0.70\pm0.17$ for B2), indicating they are gravitationally bound and likely to collapse to form a protobinary.

\end{enumerate}

Our results highlight the critical role of deuterated molecular ions in diagnosing both the physical conditions and evolutionary status of prestellar cores.


\begin{acknowledgments}
The authors thank the referees for providing thoughtful and constructive feedback that helped to improve this manuscript.
We warmly thank Chang Won Lee and Shinyoung Kim (Korea Astronomy and Space Science Institute, Republic of Korea) for providing their TRAO \NtHp(1--0) data, and Pei-Ying Hsieh (National Astronomical Observatory of Japan, Japan) for helpful discussions on single-dish data combination.
S.J.L. and S.Y.L. acknowledge grants from the National Science
and Technology Council of Taiwan (NSTC 112-2112-M-001-060-, 113-2112-M-001-004-, and 114-2112-M-001-035-MY3).
D.S. acknowledges the support from Ramanujan Fellowship (ANRF, RJF/2021/000116) and PRL, India.
S.Y.H. acknowledges the support from the Acadamia Sinica of Taiwan (grant No. AS-PD-1142-M02-2) and the National Science and Technology Council of Taiwan (NSTC 112-2112-M-001- 039-MY3). 
This work used high-performance computing facilities operated by the
Center for Informatics and Computation in Astronomy (CICA) at National
Tsing Hua University. This equipment was funded by the Ministry of
Education of Taiwan, the National Science and Technology Council of 
Taiwan, and National Tsing Hua University.
This paper makes use of the following ALMA data: 
ADS/JAO.ALMA \#2022.1.01603.S, 
ADS/JAO.ALMA \#2021.1.00546.S, 
and ADS/JAO.ALMA \#2016.1.01338.S.
ALMA is a partnership of ESO (representing its member states), NSF (USA) and NINS (Japan), together with NRC (Canada), NSTC and ASIAA (Taiwan), and KASI (Republic of Korea), in cooperation with the Republic of Chile. The Joint ALMA Observatory is operated by ESO, AUI/NRAO and NAOJ. 
The James Clerk Maxwell Telescope is operated by the East Asian Observatory on behalf of The National Astronomical Observatory of Japan; Academia Sinica Institute of Astronomy and Astrophysics; the Korea Astronomy and Space Science Institute; the National Astronomical Research Institute of Thailand; Center for Astronomical Mega-Science (as well as the National Key R\&D Program of China with No. 2017YFA0402700). Additional funding support is provided by the Science and Technology Facilities Council of the United Kingdom and participating universities and organizations in the United Kingdom and Canada.
Additional funds for the construction of SCUBA-2 were provided by the Canada Foundation for Innovation. 
The authors wish to recognize and acknowledge the very significant cultural role and reverence that the summit of Maunakea has always had within the indigenous Hawaiian community.  We are most fortunate to have the opportunity to conduct observations from this mountain.
This research used the facilities of the Canadian Astronomy Data Centre operated by the National Research Council of Canada with the support of the Canadian Space Agency.
\end{acknowledgments}

%

\vspace{5mm}
\facilities{Atacama Large Millimeter/submillimeter Array (ALMA), James Clerk Maxwell Telescope (JCMT).}


\software{Astropy \citep{astropy1, astropy2},
          CASA \citep{casa2022},
          CARTA \citep{carta2021},
          SciPy \citep{scipy},
          Starlink \citep{Currie14}}



\appendix

\setcounter{table}{0}
\setcounter{figure}{0}
\setcounter{equation}{0}
\renewcommand{\thetable}{A\arabic{table}}
\renewcommand{\thefigure}{A\arabic{figure}}
\section{Correlator Setups and Calibrators}\label{app:other_data}

The ALMA Band 7 and Band 3 receivers were used to capture data with four spectral windows (SPWs) each, 
as summarized in \autoref{tab:band7} and \autoref{tab:band3}, respectively.
We also list the channel spacing and spectral resolution in kHz in the tables above, and the corresponding values in \kms for the \oHtDp, \NtHp, and \NtDp lines are provided in \autoref{tab:obs}.
The spectral resolution values are provided for completeness and are used for velocity dispersion analysis in Section \ref{sec:ana_gas}.
The spectral resolution, defined as the full width at half maximum (FWHM) of the spectral response function, differs from the channel spacing in \autoref{tab:band7} and \autoref{tab:band3} due to the effects of the applied weighting function and online channel averaging\footnote{https://almascience.nrao.edu/documents-and-tools/cycle9/alma-technical-handbook}\footnote{https://safe.nrao.edu/wiki/pub/Main/ALMAWindowFunctions/Note\_on\_Spectral\_Response.pdf}. 
By default, ALMA correlators apply a Hanning weighting function to the data, 
resulting in the spectral resolution being twice the channel spacing when no online channel averaging ($N_{\rm avg}=1$) is applied prior to data storage.
When $N_{\rm avg}>1$, the output channels become more independent.
The default $N_{\rm avg}=2$ is adopted for SPWs targeting molecular lines, 
except for \NtHp(1--0) and HNC (1--0),
for which $N_{\rm avg}=1$ is used to 
preserve the native spectral resolution useful for gas-kinematic analysis. Results will be presented in a future publication.

\begin{deluxetable}{clrcrr}
\caption{Band 7 Correlator Setup}
\label{tab:band7}
\tablehead{
\colhead{\shortstack{Central frequency \\ of the spectral window}} & \colhead{Main tracer} & \colhead{Bandwidth} & \colhead{\shortstack{Online channel\\ averaging factors ($N_{\rm avg}$)}} & \colhead{\shortstack{Channel\\ spacing}} & \colhead{\shortstack{Spectral\\ resolution}} \\
\colhead{(GHz)} & \colhead{} & \colhead{(MHz)} & \colhead{} & \colhead{(kHz)} & \colhead{(kHz)}
}
\colnumbers
\startdata
357.973 & Continuum         & 1875     & 1  & 15625    & 31250 \\
360.163 & \DCOp(5--4)       & 234.375  & 2 & 122.070   & 141.113 \\
372.414 & \oHtDp(\oHtDpgrd)  & 234.375  & 2 & 122.070   & 141.113 \\
372.665 & \NtHp(4--3)       & 234.375  & 2 & 122.070   & 141.113 \\
\enddata
\end{deluxetable}

\begin{deluxetable}{clrcrr}
\caption{Band 3 Correlator Setup}
\label{tab:band3}
\tablehead{
\colhead{\shortstack{Central frequency \\ of the spectral window}} & \colhead{Main tracer} & \colhead{Bandwidth} & \colhead{\shortstack{Online channel\\ averaging factor ($N_{\rm avg}$)}} & \colhead{\shortstack{Channel\\ spacing}} & \colhead{\shortstack{Spectral\\ resolution}} \\
\colhead{(GHz)} & \colhead{} & \colhead{(MHz)} & \colhead{} & \colhead{(kHz)} & \colhead{(kHz)}
}
\colnumbers
\startdata
90.650 & HNC (1--0)         & 58.594   & 1 & 15.259  & 30.518 \\
90.673 & CCS (7$_7$--6$_6$) & 58.594   & 2 & 30.518  & 35.278 \\
91.486 & Continuum          & 937.500  & 4 & 976.562 & 976.563 \\
93.160 & \NtHp(1--0)        & 58.594   & 1 & 15.259  & 30.518 \\
\enddata
\end{deluxetable}

\setcounter{table}{0}
\setcounter{figure}{0}
\setcounter{equation}{0}
\renewcommand{\thetable}{B\arabic{table}}
\renewcommand{\thefigure}{B\arabic{figure}}
\section{Derivation of Column Densities for \oHtDp and \Ht}\label{app:colden}

We describe the derivation of the column densities of o-\HtDp and \Ht. 
For the o-\HtDp($J_{\rm K_a,K_c}$ = \oHtDpgrd) transition, 
the line optical depth per channel $\tau_\nu$ at each pixel is calculated from the brightness temperature $T_{\rm b}$, assuming a constant excitation temperature $T_{\rm ex}$ along the line of sight, following \citet{Ulich76}:
\begin{equation}
\tau_\nu=-\log(1-\frac{T_{\rm b}/\phi}{J_\nu(T_{\rm ex})-J_\nu(T_{\rm bg})}), \label{equ:tau}
\end{equation}
where $J_\nu(T)=(h\nu/k)/[\exp(h\nu/(kT))-1]$ is the effective radiation temperature at the rest frequency $\nu$ of the \oHtDpgrd transition, $T_{\rm bg}=2.725$~K is the cosmic microwave background temperature, and $\phi$ is the filling factor, assumed to be 1. 
The column density of o-\HtDp is then derived under the LTE condition \citep[e.g.,][]{Caselli08, Mangum15}:
\begin{equation}
N(\mathrm{o\text{-}H_2D^+})=\frac{8\pi \nu^3}{c^3A_{\rm ul}}\frac{Q(T_{\rm ex})}{g_{\rm u}\exp(-E_{\rm u}/(k_{\rm B}T_{\rm ex}))}\frac{\int\tau_\nu dv}{\exp(h\nu/(k_{\rm B}T_{\rm ex}))-1},
\end{equation}
where $A_{\rm ul}$ is the Einstein coefficient, $g_{\rm u}$ is the upper-level degeneracy, $E_{\rm u}$ is the upper-level energy, $Q(T_{\rm ex})$ is the partition function,
$k_{\rm B}$ is the Boltzmann constant, and $h$ is the Planck constant.
Values for $A_{\rm ul}$, $g_{\rm u}$, and $E_{\rm u}$ for $J_{\rm K_a,K_c}$=\oHtDpgrd, as well as the ortho-form partition function for transitions available up to $J=4$ and $K_{\rm a}=3$, are obtained from the CDMS database \citep{Muller05CDMS}.

For the 820~\micron dust continuum, the column density of \Ht is derived assuming optically thin emission and a constant dust temperature $T_{\rm d}$ along the line of sight,
following \citet{Kauffmann08}:
\begin{equation}
N(\mathrm{H_2})=\frac{S_\nu}{\Omega_{\rm bm} \mu_{\rm H_2} m_{\rm H} \kappa_\nu r_{\rm d/g} B_\nu(T_{\rm d})},
\end{equation}
where $S_\nu$ is the flux density per beam, $\Omega_{\rm bm}$ is the beam solid angle, 
$\mu_{\rm H_2}=2.8$ is the molecular weight per hydrogen molecule, 
$m_{\rm H}$ is the mass of a hydrogen atom,
$\kappa_\nu=2.27$~cm$^2$~g$^{-1}$ is the dust absorption coefficient at 820~\micron adopted from \citet{Ossenkopf94} for thick icy dust at $10^6$~cm$^{-3}$, $r_{\rm d/g}=0.01$ is the dust-to-gas mass ratio, and $B_\nu(T_{\rm d})$ is the Planck function at frequency $\nu$ and dust temperature $T_{\rm d}$. 

\setcounter{table}{0}
\setcounter{figure}{0}
\setcounter{equation}{0}
\renewcommand{\thetable}{C\arabic{table}}
\renewcommand{\thefigure}{C\arabic{figure}}
\section{Velocity Dispersion Calculation}\label{app:dispersion}
We consider four components contributing to the measured o-\HtDp velocity dispersion, $\sigma_v$(o-\HtDp) by thermal broadening ($\sigma_{\rm th}$), nonthermal broadening ($\sigma_{\rm nt}$), 
unresolved hyperfine structure ($\sigma_{\rm hfs}$),
and instrumental channel broadening ($\sigma_{\rm ch}$) with
\begin{equation}
\sigma_v^2(\mathrm{o\text{-}H_2D^+}) = \sigma_{\rm th}^2 + \sigma_{\rm nt}^2 + \sigma_{\rm hfs}^2 + \sigma_{\rm ch}^2, \label{equ:dis}
\end{equation}
where the intrinsic velocity dispersion is 
\begin{equation}
\sigma_{\rm intr}^2= \sigma_{\rm th}^2 + \sigma_{\rm nt}^2 + \sigma_{\rm hfs}^2.
\end{equation}

The thermal velocity dispersion is 
\begin{equation}
\sigma_{\rm th}^2=k_{\rm B}T_{\rm k}/m, 
\end{equation}
where $T_{\rm k}$ is the kinetic temperature, and $m=4$~amu is the molecular mass of o-\HtDp. 
The resulting $\sigma_{\rm th}$ is 0.129~\kms at $T_{\rm k}=8$~K, and 0.102~\kms at $T_{\rm k}=5$~K.
The isothermal sound speed is 
\begin{equation}
c_s^2=k_{\rm B}T_{\rm k}/(\mu m_{\rm H}), \label{equ:cs}
\end{equation}
where $\mu = 2.33$ is the mean molecular weight per particle.
The resulting $c_s$ is 0.169~\kms at $T_{\rm k}=8$~K, and 0.134~\kms at $T_{\rm k}=5$~K. The nonthermal velocity dispersion is related to the Mach number $\mathcal{M}$ via
\begin{equation}
\sigma_{\rm nt}=\mathcal{M}c_s.
\end{equation}

We include the contribution from unresolved hyperfine structure (HFS) as an additional variance component $\sigma_{\rm hfs}^2$.
For hyperfine components with velocity offsets $v_i$ and normalized relative strengths $w_i$ ($\sum_i w_i=1$), the contribution is given by the second central moment of $v_i$,
\begin{equation}
\sigma_{\rm hfs}^2=\sum_i (v_i-\bar{v})^2w_i, \quad \bar{v}=\sum_i v_i w_i.\label{equ:RMS_hyp}
\end{equation}
Using the o-\HtDp(\oHtDpgrd) hyperfine offsets and strengths from \citet{Jensen97}, we obtain $\sigma_{\rm hfs}=0.079$~\kms. 
The normalized hyperfine components are also shown in black beneath the B1 spectrum on \autoref{fig:mom}(c) for reference.
When thermal and nonthermal broadening are small enough that parts of the hyperfine pattern are marginally resolved, approximating the HFS contribution with 
$\sigma_{\rm hfs}$ and fitting a single Gaussian provides an upper bound on the FWHM.
For the \oHtDpgrd line, the Gaussian FWHM exceeds the forward-modeled hyperfine profile by 5\% in the purely thermal 5~K case (where $\sigma_{\rm th}/\sigma_{\rm hfs}=1.3$), and by only 1.5\% in the purely thermal 8~K case (where $\sigma_{\rm th}/\sigma_{\rm hfs}=1.64$).

Instrumental channel broadening can be estimated by treating the ALMA spectral response (see \autoref{app:other_data}) as a Gaussian with FWHM $\delta v_{\rm res}$, so that
\begin{equation}
\sigma_{\rm ch} \approx \delta v_{\rm res}/\sqrt{8\log 2}.
\end{equation}
However, \citet{Koch18} showed this correction is valid only when the intrinsic velocity dispersion is well sampled by the spectral channels. If 
$\sigma_{\rm intr} < 2\delta v_{\rm ch} $ (where $\delta v_{\rm ch}$ is the channel spacing),
the Gaussian approximation $\delta v_{\rm res}/\sqrt{8\log 2}$ 
tends to overestimate $\sigma_{\rm ch}$. 
In this undersampled regime, $\sigma_{\rm ch}$ should be obtained by forward modeling of the spectral response.
We therefore set $\sigma_{\rm ch}=0$ as a conservative workaround in place of forward modeling with the caveat that the measured $\sigma_v$ may be larger than $\sigma_{\rm intr}$ by $\gtrsim 5\%$ (and by $\gtrsim 10\%$ when $\sigma_{\rm intr}<\delta v_{\rm ch}$) according to \citet{Koch18}.
For our o-\HtDp data, $\delta v_{\rm res}=0.114$~\kms and $\delta v_{\rm ch}=0.098$~\kms (see \autoref{tab:obs}). We adopt
\begin{equation}
\sigma_{\rm ch} =
\begin{cases}
\delta v_{\rm res}/\sqrt{8\log 2} = 0.048~\text{\kms,} & \text{if } \sigma_{\rm intr}\ge 2\delta v_{\rm ch} =0.196~\text{\kms,}\\[4pt]
0~\text{\kms,} & \text{otherwise}.
\end{cases}
\end{equation}

\autoref{fig:mom}(d) shows three reference $\sigma_v(\mathrm{o\text{-}H_2D^+})$ cases. Only the $\mathcal{M}=1$ case ($\sigma_{v}=0.232$~\kms, 8~K) includes a nonzero $\sigma_{\rm ch}$ term, because $\sigma_{\rm intr}=0.227~\text{\kms}\ge 2\delta v_{\rm ch}$. The other two cases yield $\sigma_v=\sigma_{\rm intr}=0.129$~\kms (thermal, 5~K), and 0.151~\kms (thermal, 8~K).
Even in the undersampled thermal cases, setting $\sigma_{\rm ch}=0$ underestimates $\sigma_v$ by $\sim$5\% \citep{Koch18}, while the single Gaussian HFS approximation overestimates it by 5\% in our narrowest case; both effects are within the fitting uncertainties, so the resulting $\sigma_v$ values remain reliable.

\setcounter{table}{0}
\setcounter{figure}{0}
\setcounter{equation}{0}
\renewcommand{\thetable}{D\arabic{table}}
\renewcommand{\thefigure}{D\arabic{figure}}
\section{Virial Parameters for Substructures B1 and B2}\label{app:mass}

\begin{figure*}[ht]
    \centering
    \includegraphics[width=0.8\textwidth]{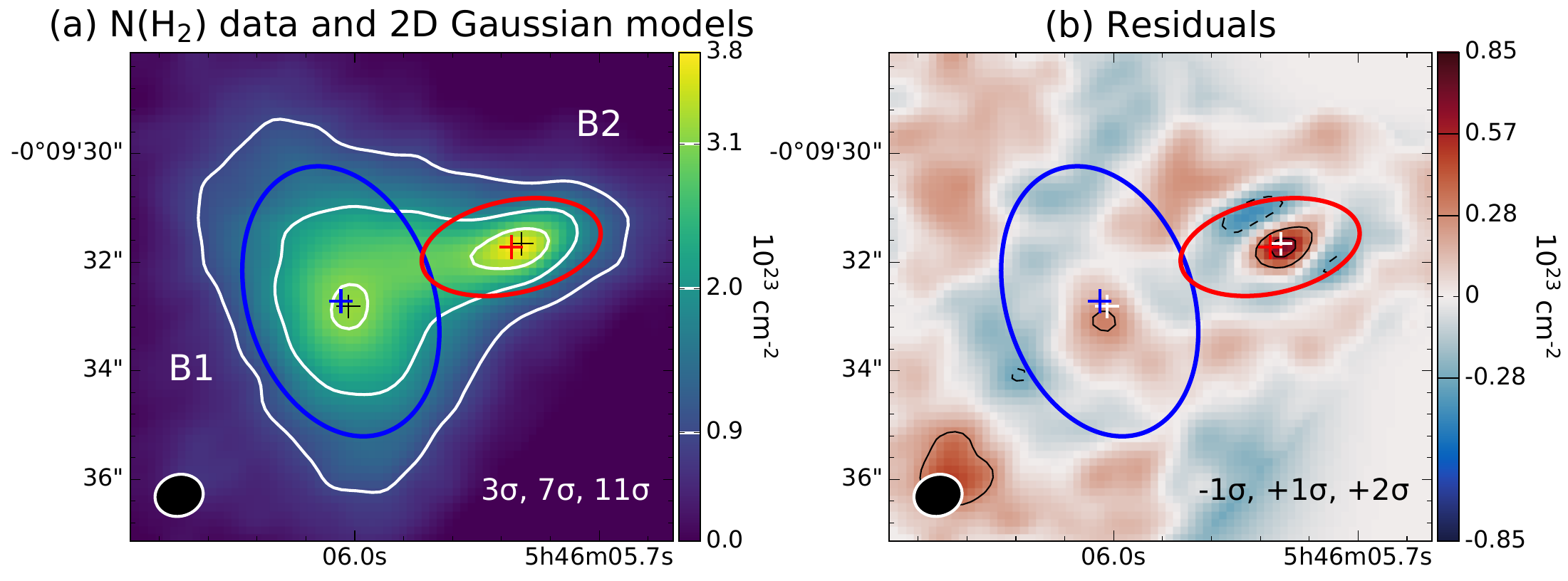}
    \caption{
    Mass derivation of \soufullname substructures B1 and B2 from our 820~\micron data.
    \textbf{(a)} $N$(\Ht) map and contours from \autoref{fig:LTE}(c) overlaid with 2D Gaussian models, where ellipses indicate the FWHMs of the major and minor axes. 
    \textbf{(b)} Residuals between the observed data and the fitted models, 
    with $N$(\Ht) contour levels at $-1\sigma$, $1\sigma$, and $2\sigma$, where $\sigma=2.8\times10^{22}$~cm$^{-2}$.
    Blue and red crosses mark the fitted Gaussian centers, while black/white crosses denote the 2D Gaussian centers derived from 1.3~mm data \citep[Table 1 from][]{Sahu21}.
    }
    \label{fig:masses}
\end{figure*}

\begin{deluxetable}{ccccccccc}
\caption{Summary of the 2D Gaussian Mass Parameters and the Virial Parameters}
\label{tab:masses}
\tablehead{
\colhead{Substructure} & \colhead{Center} & \colhead{$N_0$(\Ht)} & \colhead{$a_{\rm FWHM}$} & \colhead{$b_{\rm FWHM}$} & \colhead{P.A.} & \colhead{Mass} & \colhead{$\sigma_{\rm tot}$} & \colhead{$\alpha$} \\
\colhead{} & \colhead{($\alpha_{\rm ICRS}$, $\delta_{\rm ICRS}$)} & \colhead{(10$^{23}$ cm$^{-2}$)} & \colhead{(arcsec.)} & \colhead{(arcsec.)} & \colhead{(degree)} & \colhead{(M$_\odot$)} & \colhead{(\kms)} & \colhead{}
}
\colnumbers
\startdata
B1 & $5^{\mathrm h}46^{\mathrm m}06\fs017$, $-0\degr09\arcmin32\farcs72$ & $2.8\pm0.2$ & $5\farcs10\pm0\farcs29$ & $3\farcs44\pm0\farcs22$ & $+18\degr\pm5\degr$ & $0.47\pm0.05$ & $0.198\pm0.005$ & $0.40\pm0.05$ \\
B2 & $5^{\mathrm h}46^{\mathrm m}05\fs808$, $-0\degr09\arcmin31\farcs72$ & $3.0\pm0.3$ & $3\farcs35\pm0\farcs34$ & $1\farcs71\pm0\farcs16$ & $-78\degr\pm5\degr$ & $0.16\pm0.03$ & $0.203\pm0.013$ & $0.70\pm0.17$ \\
\enddata
\tablecomments{Columns (2)--(6): 2D Gaussian-fitted parameters. The P.A. (major axis position angle) is measured counterclockwise from the north.}
\end{deluxetable}

We first derive the masses of B1 and B2.
We assume a Gaussian mass distribution for the substructures and simultaneously fit the $N$(\Ht) map with two 2D Gaussian components, as shown in \autoref{fig:masses}. 
The residuals being below $3\sigma$ demonstrate that the 2D Gaussian models provide a good fit to the data.
The Gaussian-fitted centers for B1 and B2 in our 820~\micron data are consistent with those derived from the 1.3mm data by \citet{Sahu21}, with offsets smaller than both our 820~\micron beam size of $\sim$0\farcs8 and the 1.3~mm beam size of $\sim$1\farcs2, suggesting that both wavelengths reliably trace the same substructures.

The mass of each substructure is then calculated by
\begin{equation}
M = \mu_{\rm H_2} m_{\rm H} N_0(\mathrm{H_2}) D^2 \Omega_{\rm 2D},
\end{equation}
where $N_0$(\Ht) is the fitted Gaussian amplitude, $D=400$~pc is the distance to \souname, and $\Omega_{\rm 2D}$ is the solid angle subtended by the fitted 2D Gaussian distribution, defined by
\begin{equation}
\Omega_{\rm 2D} = 2\pi \frac{a_{\rm FWHM} b_{\rm FWHM}}{8 \log 2},
\end{equation}
where $a_{\rm FWHM}$ and $b_{\rm FWHM}$ are the FWHMs along the major and minor axes, respectively. The fitted parameters and the derived masses are summarized in \autoref{tab:masses}.

We calculate the virial mass by considering gravitational and kinetic energy, assuming a uniform sphere, with
\begin{equation}
M_{\rm vir}=5\sigma_{\rm tot}^2R_{\rm eff}/G,
\end{equation}
where $\sigma_{\rm tot}$ is the total dispersion, and $R_{\rm eff}$ is the geometric mean of $a_{\rm FWHM}$ and $b_{\rm FWHM}$.
Using \autoref{equ:dis} and \autoref{equ:cs}, we compute the total dispersion by
\begin{equation}
\sigma_{\rm tot}^2=c_s^2 + \sigma_{\rm nt}^2=k_{\rm B}T_{\rm k}/(\mu m_{\rm H}) +\sigma_v^2(\mathrm{o\text{-}H_2D^+})-\sigma_{\rm th}^2-\sigma_{\rm hfs}^2-\sigma_{\rm ch}^2.
\end{equation}
The virial parameter is defined as
\begin{equation}
\alpha=M_{\rm vir}/M,
\end{equation}
with $\alpha=1$ indicating virial equilibrium, and $\alpha<2$ commonly used as a criterion of gravitational binding \citep{Myers21}.
Using our measured o-\HtDp(\oHtDpgrd) dispersion of $0.183\pm0.005$~\kms for B1 and $0.188\pm0.014$~\kms for B2 (see \autoref{fig:mom}(c)), we compute the total dispersions at $T_{\rm k}=8$~K and derive the corresponding virial parameters, as summarized in \autoref{tab:masses}.



\bibliography{main.bib}
\bibliographystyle{aasjournal}



\end{document}